%% file: fgcs_enginecl.tex
\documentclass[5p]{elsarticle}
\usepackage{makeidx}  
\usepackage{wrapfig} 
\usepackage{multirow}
\input{insbox}
\usepackage{moresize}

\usepackage[hyphens]{url}

\usepackage{tikz}
\usetikzlibrary{calc,arrows}

\usepackage{xcolor}
\usepackage{colortbl}

\usepackage{tabularx}
\usepackage{array}

\usepackage{stfloats} 
\usepackage{setspace}

\usepackage{ulem}
\definecolor{light-gray}{gray}{0.95}
\definecolor{MyDarkGreen}{rgb}{0,0.8,0.0}
\definecolor{MyDarkBlue}{rgb}{0,0,0.8}
\definecolor{MyDarkRed}{rgb}{0.6,0,0.0}
\definecolor{MyOrange}{rgb}{0.6,0,0.0}

\usepackage{multicol} 
\usepackage{relsize}
\usepackage{xspace}
\PassOptionsToPackage{hyphens}{url}\usepackage{hyperref}

\newcommand\Cpp{C\nolinebreak[4]\hspace{-.05em}\raisebox{.4ex}{\relsize{-3}{\textbf{++}}}}

\usepackage{fancyhdr}
																	     \fancypagestyle{pprintTitle}{%
\chead{Accepted in Future Generation Computer Systems: \url{https://doi.org/10.1016/j.future.2020.02.016}}%
																					     \lfoot{}\cfoot{}\rfoot{}
																							     
																									     }
\usepackage[frozencache,cachedir=.]{minted}

\usepackage[most,minted]{tcolorbox}
\tcbuselibrary{minted,skins}
\usepackage{etoolbox}

\newtcblisting{snippet}[2][]{
    listing engine=minted,
    colback=light-gray,
    colframe=white,
    listing only,
    minted style=default,
    minted language=#2,
    minted options={linenos=true,numbersep=2mm,fontsize=\small,#1},
    left=18pt,
    enhanced,
    overlay={\begin{tcbclipinterior}\fill[black!15] (frame.south west)
            rectangle ([xshift=6mm]frame.north west);\end{tcbclipinterior}}
}
\usepackage{graphicx}
\usepackage{subcaption}

\usepackage{caption} 

\usepackage{amssymb}
\usepackage{amsmath}
\usepackage{mathtools}
\DeclarePairedDelimiter{\floor}{\lfloor}{\rfloor}
\renewcommand{\arraystretch}{1.1}

\usepackage{hyperref} 

\hypersetup{
bookmarks=true,         
unicode=false,          
pdftitle={EngineCL: Usability and Performance in Heterogeneous Computing. [Nozal et al. 2019]},
pdfauthor={Raul Nozal, Jose Luis Bosque and Ramon Beivide},
pdfkeywords={
Heterogeneous Computing,
Effortless co-execution,
Usability,
Performance portability,
OpenCL,
Scheduling,
Load balancing,
Productivity,
API
},
pdfsubject={System Software},
pdfproducer={LaTeX,TeX},
pdfcreator={xelatex}
}

\begin{document}


%
%
\title{EngineCL: Usability and Performance in Heterogeneous Computing}



\author[cant]{Ra\'ul Nozal%
  \corref{cor1}%
  }%
  \ead{raul.nozal@unican.es}
\author[cant]{Jose Luis Bosque%
  }%
  \ead{joseluis.bosque@unican.es}
\author[cant]{Ramon Beivide
  }%
  \ead{ramon.beivide@unican.es}
  \cortext[cor1]{Corresponding author}

\address[cant]{Computer Science and Electronics Department, Universidad de Cantabria, Spain}

%
%
%

\input{abstract.tex}

\maketitle              



\input{intro.tex}

\input{background.tex}

\input{motivation.tex}

\input{design.tex}

\input{implementation.tex}

\input{api.tex}

\input{methodology.tex}

\input{validation.tex}

\input{related.tex}

\input{conclusions.tex}

\section*{Acknowledgement}
This work has been supported by the Spanish Ministry of Education (FPU16/ 03299 grant), the Spanish Science and Technology Commission (TIN2016-76635-C2-2-R) and the European HiPEAC Network of Excellence.

\input{biblio.tex}

\input{bio.tex}



\end{document}

%% file: abstract.tex
\begin{abstract}
Heterogeneous systems
have become one of the most common architectures today, thanks to their excellent performance and energy consumption. However, due to their heterogeneity they are very complex to program and even more to achieve performance portability on different devices. This paper presents EngineCL, a new OpenCL-based runtime system that outstandingly simplifies the co-execution of a single massive data-parallel kernel on all the devices of a heterogeneous system. It performs a set of low level tasks regarding the management of devices, their disjoint memory spaces and scheduling the workload between the system devices while providing a layered API. EngineCL has been validated in two compute nodes (HPC and commodity system), that combine six devices with different architectures. Experimental results show that it has excellent usability compared with OpenCL; a maximum 2.8\% of overhead compared to the native version under loads of less than a second of execution and a tendency towards zero for longer execution times; and it can reach an average efficiency of 0.89 when balancing the load.
\end{abstract}


\begin{keyword}
Heterogeneous Computing \sep
Usability \sep
Performance portability \sep
OpenCL \sep
Parallel Programming \sep
Scheduling \sep
Load balancing \sep
Productivity \sep
API
\end{keyword}

%% file: intro.tex
\section{Introduction}
\label{sec:intro}

The emergence of heterogeneous systems is one of the most important milestones in parallel computing in recent years \cite{Zahran:2016}. A heterogeneous system is composed of general purpose CPUs and specific purpose hardware accelerators, such as GPUs, Xeon Phi, FPGAs or 
TPUs.
Under this concept, a wide range of systems are included, from powerful computing nodes capable of executing teraflops \cite{NVIDIADGX}, to integrated CPU and GPU chips \cite{iGPU}. This architecture allows, not only to significantly increase the computing power, but also to improve their energy efficiency.

However, this architecture also presents a series of challenges, among which the complexity of its programming and the performance portability stand out. In this sense, the Open Computing Language (OpenCL) has been developed as an API that extends the C/\Cpp{} programming languages for heterogeneous systems \cite{opencl:Gaster:2013}. OpenCL provides low abstraction level that forces the programmer to know the system in detail, determining the architecture of the devices, managing the host-device communication, understanding the distributed address memory space and explicitly partitioning the data among the devices, transferring the input data and collecting the results generated in each device. The management of these aspects greatly complicates programming, which turns into an error-prone process, significantly reducing the productivity \cite{Maat:2009}.

On the other hand, OpenCL follows the Host-Device programming model. Usually the host (CPU) offloads a very time-consuming function (kernel) to execute in one of the devices.  
If there is only one function that can be executed at any given time, both the CPU and the rest of the devices are idle waiting for completion, 
consuming energy and wasting their computing capacity.
To overcome this problem, it is necessary to propose a paradigm change in the programming model and to encourage data-parallelism. This is achieved through \emph{co-execution}, defined as the collaboration of all the devices in the system (including the CPU) to execute a single massive data-parallel kernel \cite{Zhang:17, Shen:16, Stafford:17}. However, it is a hard task for the programmer and needs to be done effortless in order to be widely used.
In this context, this paper presents {\em EngineCL}, a new OpenCL-based \Cpp{} runtime API that significantly improves the usability of the heterogeneous systems without any loss of performance\footnote{The EngineCL code is available at: \url{https://github.com/EngineCL/EngineCL}}. It accomplishes complex operations transparently for the programmer, such as discovery of platforms and devices, data management, load balancing and robustness throughout a set of efficient techniques. EngineCL follows Architectural Principles with known Design Patterns to strengthen the flexibility in the face of changes. 
The runtime manages a single data-parallel kernel among all the devices in the heterogeneous system. Its modular architecture allows to easily incorporate different schedulers to distribute the workload
, such as those included in this work: static, dynamic and HGuided.

EngineCL has been validated both in terms of usability and performance, using two very different heterogeneous systems. The first one resembles a High Performance Computing node and it is composed of three different architectures: multi-core CPU, GPU and Xeon Phi. The second one is an example of a desktop computing node and it is made up of a multi-core CPU with an integrated GPU (iGPU) and one commodity GPU. Regarding usability, eight metrics have been used, achieving excellent results in all of them. In terms of performance, the runtime overhead compared with OpenCL is on average around 1\% when using a single device. Finally, co-execution when using the HGuided scheduler yields average efficiencies of 0.89 for the HPC node and 0.82 for the Desktop node, using regular and irregular applications.


The main contributions of this paper are the following:
\begin{itemize}
\item Presents EngineCL, a runtime that simplifies the programming of data-parallel application on a heterogeneous system.
\item Proposes a high-level layered API, focusing on maintainability, ease of use and flexibility.
\item EngineCL allows effortless co-execution, squeezing the performance out of all the devices in the system and ensuring performance portability.
\end{itemize}

The rest of this paper is organized as follows. Section \ref{sec:programming} explains the basic concepts of OpenCL programming for heterogeneous systems, useful for the rest of the paper. Next, Section \ref{sec:challenges} shows the most relevant challenges of the heterogeneous co-execution. Then \ref{sec:Design} and \ref{sec:Implementation} describe the design and implementation of EngineCL, respectively. Section \ref{sec:API} presents two examples of how to use the API. The methodology used for the validation is in Section \ref{sec:Methodology}, while the experimental results are shown in Section \ref{sec:Validation}. Finally, Section \ref{sec:Related} explains similar works while Section \ref{sec:Conclusions}, the most important conclusions and future work are presented.

%% file: background.tex

\section{Programming Heterogeneous Systems}
\label{sec:programming}

A heterogeneous system is made up of a multi-core CPU, and one or more hardware accelerators, such as GPGPUs, FPGAs or Tensor Processing Units (TPUs). Each of these devices are developed to optimize a particular type of application, so they have a very different architecture, computing capacity, and therefore, APIs.

In this context, OpenCL entails an effort to improve programmability and code portability between different heterogeneous systems \cite{opencl:Gaster:2013}. Roughly put, it consists of an extension to C/C++ that allows programmers to shift parts of their code to the accelerators, introducing the Host-Device programming model, usually computing one device at a given time (eg. GPU). OpenCL supports all of the above devices, if the vendor provides the appropriate driver. Since EngineCL is based on OpenCL, a rough description of the framework is necessary.

To enable the execution of code in such a variety of architectures, OpenCL defines the notion of \emph{context}. Which is a set of OpenCL-capable devices from the same manufacturer that are allowed a certain degree of data sharing. Each device comprises a set of \emph{compute units}, which are an abstraction of the minimum element of the device that can execute work. The mapping of compute units to actual hardware components varies between architectures. 
The code executed on the devices is encapsulated in data-parallel functions
which are known as \emph{kernels}. When one is offloaded to a device, OpenCL launches multiple instances of the kernel, each with a different portion of the data, under the 
SIMT paradigm. Each instance is called a \emph{work-item}. The programmer can decide how many items are launched by setting a parameter called \emph{global work size}. Work-items are launched in teams so they can cooperate and synchronize with each other. OpenCL ensures that the work-items of each team, or \emph{work-group}, are launched simultaneously in the same compute unit. 
Work-group size can be defined through the \emph{local work size} parameter.

Accelerators usually have a differentiated memory address space on which they can perform their computations. For this reason, kernel launches must be preceded by an input data copy phase, from the main memory to the device memory, and followed by another in the opposite direction for the results. For these operations OpenCL uses the concept of \emph{buffers}, which are a host representation of the memory of the devices in a context. These copy phases must be explicitly instructed by the programmer, which constitutes a tedious and error-prone task.

Therefore, programming a heterogeneous system can be an arduous task. Despite the increased portability OpenCL offers, the programmers still must know the architecture of the system in detail and are responsible for the time-consuming task of adapting system management and load balancing to the actual underlying system. Therefore, if optimum performance or energy efficiency is sought then the effort required is significantly higher.

%% file: motivation.tex
\section{Challenges of Heterogeneous Co-execution}
\label{sec:challenges}

The main objective of this paper is to simplify \emph{heterogeneous co-execution}, the co-execution of the a single massive data-parallel kernel, in a set of devices with different architecture and computing capacity. This objective plans a series of challenges that will be addressed and solved, which can be grouped under three fundamental concepts: abstraction, performance portability and usability.

\paragraph{Challenge 1: Abstraction} OpenCL leaves to the programmer many low-level operations that require a thorough knowledge of the underlying architecture of the heterogeneous system. Thus, the programmer is burdened with discovering the available platforms and devices, defining buffers and distributing data among all devices, launching the execution of kernels, as well as collecting partial results and organizing them properly. All this greatly complicates the programming of heterogeneous systems, reducing productivity and making it very prone to errors. Specifically, data management is a very complex aspect, since in general the devices have separate memories. Therefore, the programmer must create and manage buffers for each device, assign a portion of the data, retrieve and organize the partial results to obtain the final result of the application.

To minimize the co-execution effort, it is necessary that the programmer is not aware of many of these details of the underlying architecture in the heterogeneous system. The solution to these problems is to offer tools that provide a higher level of abstraction and take care of all these tasks. 

\paragraph{Challenge 2: Performance Portability} OpenCL solves code portability, i.e. the same kernel can run on different heterogeneous systems. 
However, performance portability goes further, by exploiting with the same code the processing resources of different heterogeneous systems. For this, two key problems have been identified: load balancing and differences in the architecture of accelerators.

The objective of load balancing is to distribute the workload among all the devices in the system proportionally to their computing power.
It is necessary to consider both the heterogeneity of the system (different devices with different computing capacity) and the behavior of kernels that can be regular or irregular. In the former, two workloads of the same size always spend the same time on the same device. 
However, in irregular kernels it is necessary to have a dynamic and adaptive algorithm that distributes the workload at runtime and can adapt to the behavior of the application.

Accelerators are hardware devices specifically designed to accelerate the execution of applications with specific properties. For instance, GPUs favor the execution of massively data-parallel kernels through using multi-threading, while FPGAs favor the execution of kernels with deeply segmented implementations. Therefore, it is often necessary to adapt the kernel to a particular device to achieve its maximum performance.
On the other hand, compiling for some devices is time consuming, therefore it is necessary to provide the binary code. However, in others, on-line compilation can provide advantages, such as parameter tuning. 
Therefore, it is important to offer the possibility of managing kernels specific to each of the devices and providing binary or source code kernels.

\paragraph{Challenge 3: Usability} OpenCL is a programming language for programming applications that run on any hardware accelerator, if the vendor provides the appropriate driver. This makes it very powerful, since it allows a portable code between very different devices. However, this power is not without complexity. Thus, the current API of OpenCL is very complex and presents a wide variety of functions with multiple and complex parameters. For example, the following functions create a memory buffer on a device and make it accessible from the host (pinned memory; error control omitted).

\vspace{-2mm}
{\small
\begin{verbatim}
buffs[0] = clCreateBuffer(context, CL_MEM_READ_ONLY
  | CL_MEM_ALLOC_HOST_PTR, sizeof(float)*nEntries,
  buffHostPtr, &err);
buffsMap[0] = clEnqueueMapBuffer(cmdQueue, buffs[0],
  false, CL_MAP_READ, offset, sizeof(float)*nEntries,
  nEventList, eventList, &event, &err);
\end{verbatim}
}
\vspace{-2mm}

On the other hand, there are currently several OpenCL specifications available (1.0, 1.1, 1.2, 2.0, 2.1, 2.2) and different devices support one or the other. The differences between these specifications are noteworthy, so that an application programmed to run on an OpenCL 2.0 cannot run in devices for previous versions. For instance, functions like {\tt enqueueNDRangeKernel}, has a different number of parameters in OpenCL 1.0 and 2.0. Another example is the support of mechanisms for the synchronization between host and device, since some specifications support asynchronous callbacks, while others force the use of blocking communication mechanisms.

Taking into account the challenges, Table \ref{tbl:model} shows an analytical model showing the growth of the density and complexity of the code in typical data-parallel OpenCL \Cpp{} programs, in relation to the number of devices, kernels and buffers used, among others. For example, if a system has three devices and the problem requires two input and one output buffers, the number of tokens to manage OpenCL buffers will increase around 135, and the lines of code to manage the OpenCL program to 18. For this reason, large redundancy problems are detected, potential sources of errors and possibilities of simplification and optimization when working with OpenCL applications. In addition, when designing and implementing load balancing algorithms, due to the complexity in handling primitives, callbacks, data partitioning, synchronization patterns and efficient multi-threading designs, a solution that facilitates the management of the heterogeneous system is necessary.

For all these reasons, it is necessary to provide programmers of heterogeneous systems with an API that will be simpler and more intuitive to use, simplifying the set of functions and their parameters. It is also necessary to have a runtime that internally manages the possible differences in configuration and functionality of the devices, regarding their OpenCL support.

\begin{table}[!b]
\begin{center}
  \caption{Analytical model that relates the lines of code (\textit{LOC}) and tokens (constants, $c$) needed in a typical data-parallel OpenCL C++ program, depending on platforms $(Pl)$, devices $(D)$, programs $(P)$, program kernels $(P_{kernels})$, program arguments $(P_{args})$, and program buffers $(P_{buffers})$.}{%
  \label{tbl:model}
\footnotesize
\begin{tabular}{p{30mm}>{\columncolor[gray]{0.95}}cc>{\columncolor[gray]{0.95}}c}
\textbf{\textbf{OpenCL Primitives}} & \textbf{\textbf{LOC}} & \textbf{\textbf{Tokens}} & \textbf{\textbf{Model}}\\
\hline

Device                          & 3 & 9 & $c\: Pl$ \\
Context                         & 1 & 3 & $c\: D$ \\
CommandQueue                    & 2 & 9 & $c\: D$ \\
Buffer                          & 3 & 15 & $c\: D\: P_{buffers}$ \\
Program                         & 6 & 21 & $c\: D\: P$ \\
Kernel                          & 2 & 8 & $c\: D\: P_{kernels}$ \\
Arg                             & 2 & 7 & $c\: D\: P_{args}\: P_{kernels}$ \\

\end{tabular}
}
\end{center}%
\end{table}


To overcome these challenges, this paper proposes EngineCL, a new runtime and API based on OpenCL that notably simplifies the programming of heterogeneous systems. The abstraction level is increased because it frees the programmer from tasks that require a specific knowledge of the underlying architecture, and that are very error prone. It ensures performance portability thanks to the integration of schedulers that successfully distribute the workload among the devices, adapting both to the heterogeneity of the system and to the behavior of the applications. And finally, the simplified and extensible API has a great impact on their usability, productivity and maintainability.

%% file: design.tex
\section{Principles of Design}
\label{sec:Design}

EngineCL is designed with many principles in mind, all around three pillars: OpenCL, Usability and Performance. The latter two are the main building blocks of the runtime. Therefore, the trade-offs have been permanently considered and analyzed when designing the runtime and its architectural principles.

\subsection{OpenCL}
\label{sec:opencl}
EngineCL is tightly coupled to OpenCL and how it works. Therefore, it is not intended to replace it, but to act as a wrapper over it. The system modules and their relationships have been defined according to the most efficient and stable patterns. Every major design decision has been benchmarked and profiled to achieve the most optimal solution in every of its parts, but mainly promoting the modules related with the data management, synchronization and API abstraction.

Based on previous knowledge when porting and using OpenCL libraries and programs, core functionalities have been designed from scratch, implementing different ideas of OpenCL workflows and comparing them with profiling tools under different architectures.
Tools like CodeXL and VTune have been studied and used to do profiling, but the lack of support for multiple devices (anything but AMD CPU/GPU or Intel GPU) and non-proprietary drivers make them neither suitable nor flexible. During the development of EngineCL custom profiling mechanisms have been used as an alternative to tools from vendors to be independent of them, and they are finally integrated into the inspector module of EngineCL.

\begin{figure}[!h]
   \centering
   \includegraphics[width=0.47\textwidth]{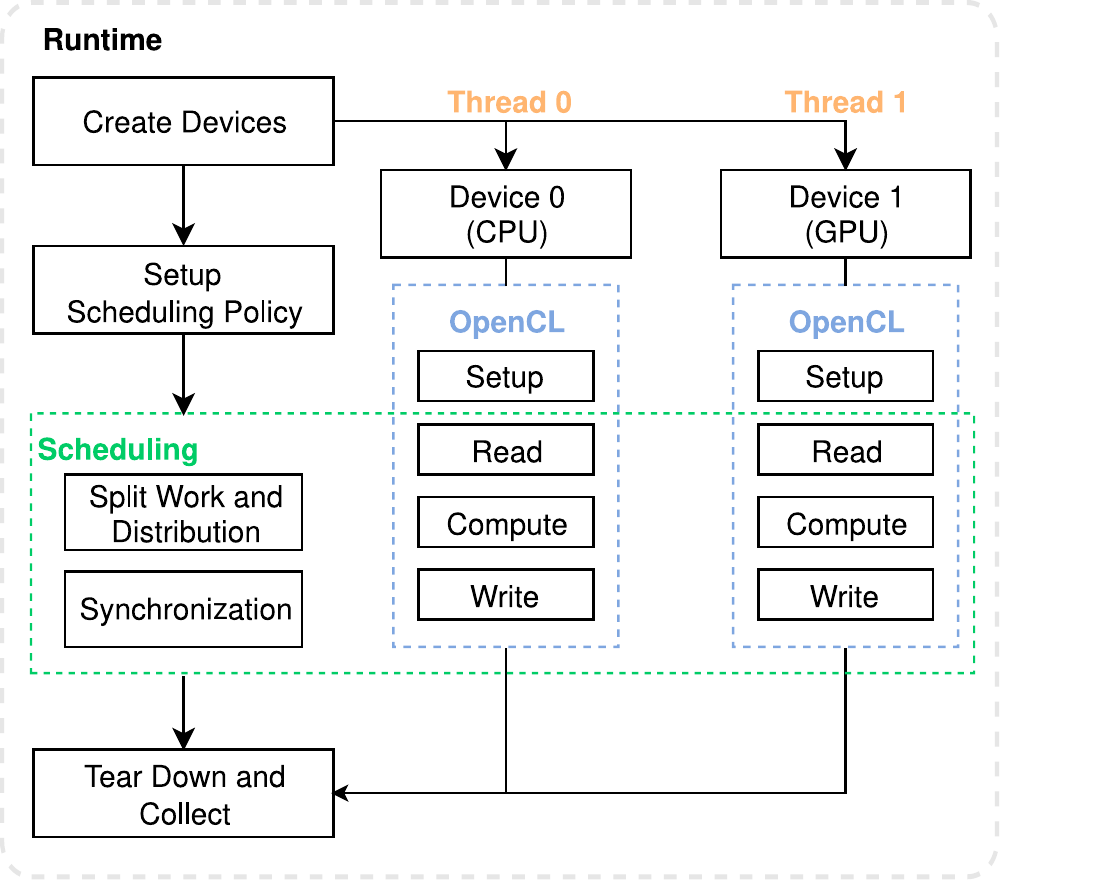}
   \hspace{-5mm}
   \caption{Scheduling view of the scheduling and work distribution. The low-level OpenCL API is encapsulated within the concept of Device, managed by a thread, being one of the core design decisions.}
   \label{fig:scheduling}
\end{figure}

Two main decisions have been applied since the very beginning: OpenCL should be isolated to improve the compatibility of the runtime and it should be managed to be easily extensible while providing the best average performance between all the available devices. The former allows high adaptation to new technologies while preserving the runtime API and its main schedulers, being slightly independent to OpenCL, but still promoting it since it is the best technology to support heterogeneous devices, as it is depicted in Figure \ref{fig:scheduling}. The latter ensures the best performance and efficiency independently of the new devices to be incorporated, solving many issues found when adapting new architectures to runtime systems. One of the core aspects to boost the performance of the system, while being able to improve the expressiveness of the schedulers, is the usage of callback mechanisms mixed with events. By doing this, it helps the drivers of the devices to optimize the enqueued operations when applicable. Although this strategy hardly complicates the internal implementation of the runtime, it allows any type of scheduling algorithm and provides asynchronous operations that boost the general efficiency for every program used with EngineCL. This is highlighted when using multiple devices. Although the API is synchronous to facilitate  its operation for common use cases, it can easily be extended to expose asynchronous behaviors to the programmer. Therefore, EngineCL has been developed along with a wide range of different architectures, always evaluating the average and peak ratios of the devices that were accessible during the development.

\begin{figure}[!b]
   \centering
   \includegraphics[width=0.49\textwidth]{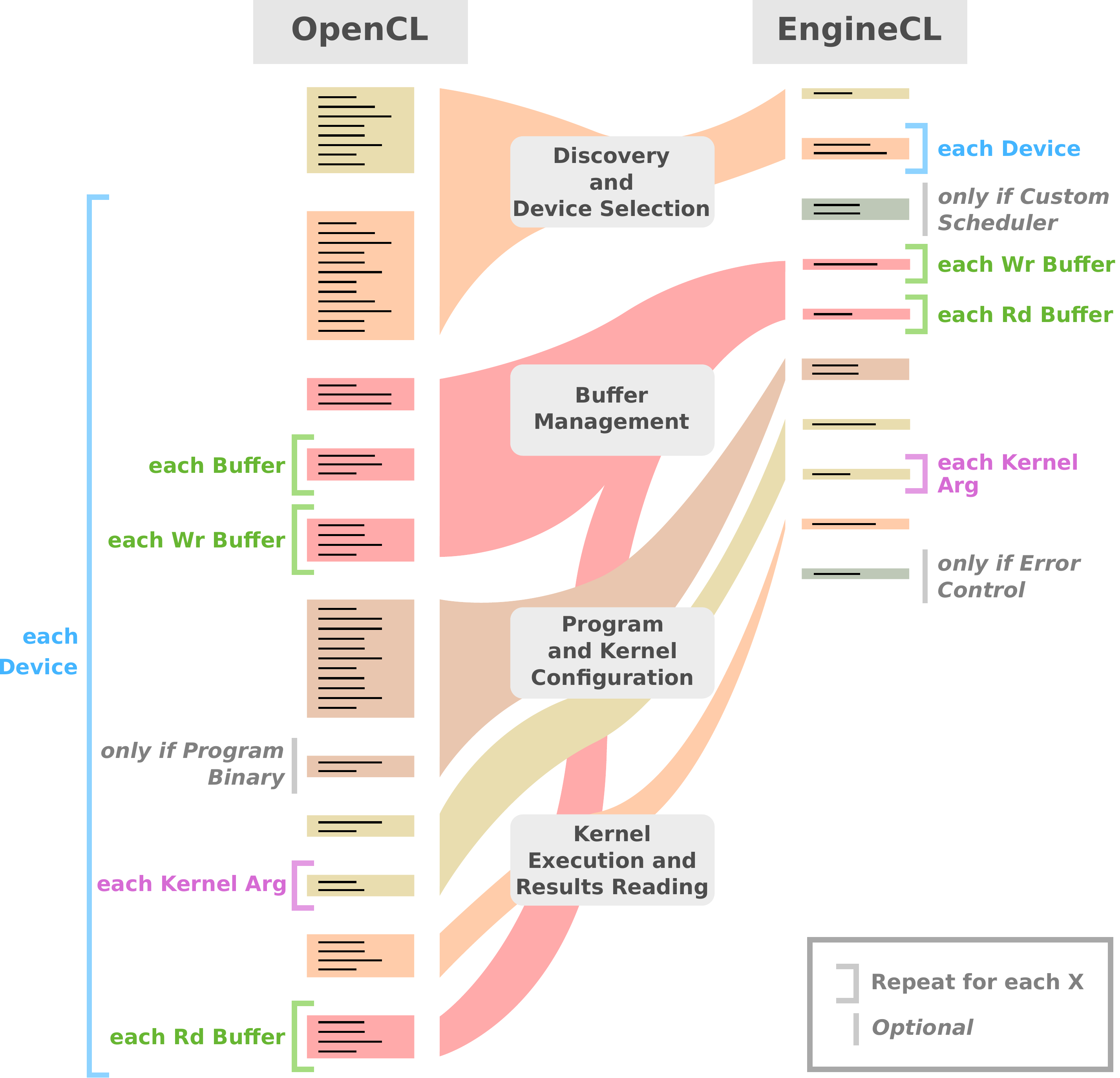}
   \caption{Overview of a generic OpenCL program and its translation to EngineCL.
     The height of every rectangle has the same proportions in lines of code as the real program. OpenCL involves more code density and repeats almost all phases per device used.}
   \label{fig:overview}
   \vspace{-5mm}
\end{figure}
\hyphenation{ma-na-ge-ment}
OpenCL presents scalability issues regarding programmability and performance when the number of devices to be used increase. Thus, it is complex to manage efficiently data structures, OpenCL primitives and call operations, as can be seen in Section \ref{sec:Validation}. Figure \ref{fig:overview} depicts a generic OpenCL program, conceptually and in density of code, compared with the EngineCL version. As the number of devices, operations and data management processes increases, the code grows quickly with OpenCL, decreasing the productivity and increasing the maintainability effort. EngineCL solves these issues by providing a runtime with a high-level API that efficiently manages all the resources of the underlying system, as will be seen in Section \ref{sec:API}.

\subsection{Architectural Principles: usability and performance}
\label{sec:architecture}



\begin{figure*}[!h]
\centering
\hspace{-3mm}
\begin{minipage}[c]{0.65\textwidth}
   \includegraphics[trim=0 8 330 0,clip,height=5.5cm,keepaspectratio]{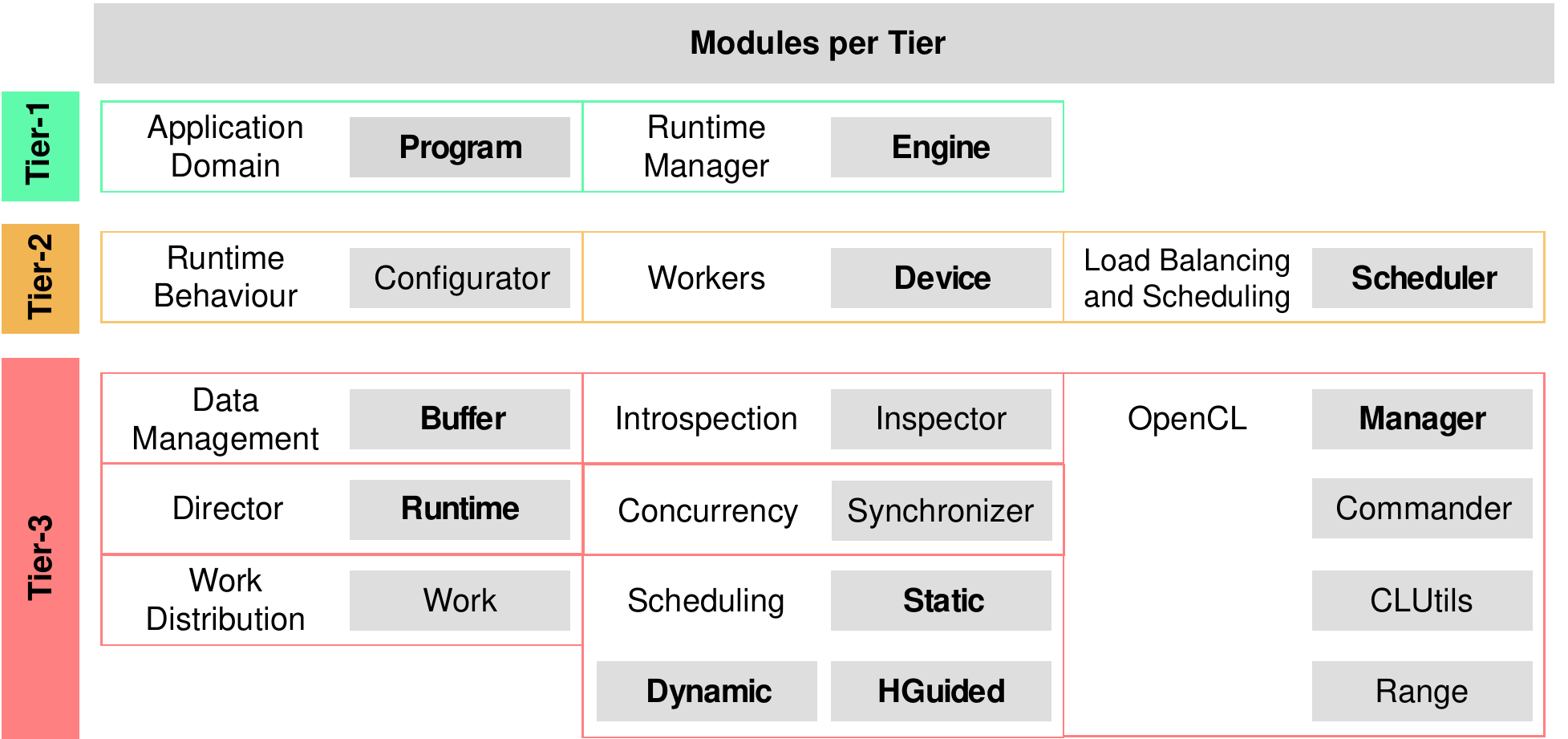}
   \vspace{-2mm}
    \caption{EngineCL building blocks: tiers, contexts and modules (main are highlighted).\\ The lower the tier, the more advanced the features (fine-grained management). Each\\ module belongs to a conceptual context (eg. HGuided to Scheduling).}\label{fig:tiers}
\end{minipage}
\hspace{-5mm}
\begin{minipage}[c]{0.34\textwidth}
   \includegraphics[trim=560 8 00 0,clip,height=5.5cm,keepaspectratio]{modules_relations_v2.pdf}
   \vspace{-6mm}
  \caption{Design Patterns of the main modules. The modules of the upper layers abstract and facilitate the use of the lower ones.}\label{fig:tiers-relations}
\end{minipage}
\vspace{-4mm}
\end{figure*}

EngineCL redefines the concept of \textit{program} to facilitate its usage and the understanding of a kernel execution. Because a program is associated with the application domain, it has data inputs and outputs, a kernel and an output pattern. The data is materialized as \Cpp{} containers (like \textit{vector}), memory regions (C pointers) and kernel arguments (POD-like types, pointers or custom types). The kernel accepts directly an OpenCL-kernel string, and the output pattern is the relation between the \textit{global work size} and the size of the output buffer written by the kernel. The default value is $1:1$, because every work-item (thread) writes to a single position in the output buffers ($\frac{1\ out\ index}{1\ work-item}$, e.g. the third work-item writes to the third index of every output buffer). The programmer guides with this relation to avoid incurring in performance-penalties trying to guess or analyzing the kernel at runtime. The programmer can easily analyze the kernel to provide the output pattern. 

It is designed to support massive data-parallel kernels, but thanks to the program abstraction the runtime will be able to orchestrate multi-kernel executions (task-parallelism), prefetching of data inputs, optimal data transfer distribution, iterative kernels and track kernel dependencies and act accordingly. Therefore, the architecture of the runtime is not constrained to a single program.

The runtime follows Architectural Principles with well-known Design Patterns to strengthen the flexibility in the face of changes. As it is depicted in Figure \ref{fig:tiers}, the runtime is layered in three tiers, and its implementation serves the following purposes: Tier-1 and Tier-2 are accessible by the programmer. The lower the Tier, the more functionalities and advanced features can be manipulated. Most programs can be implemented in EngineCL with just the Tier-1, by using the \textit{EngineCL} and \textit{Program} classes. The Tier-2 should be accessed if the programmer wants to select a specific \textit{Device} and provide a specialized kernel, use the \textit{Configurator} to obtain statistics and optimize the internal behavior of the runtime or set options for the \textit{Scheduler}. Tier-3 contains the hidden inner parts of the runtime that allows a flexible system regarding memory management, pluggable schedulers, work distribution, high concurrency and OpenCL encapsulation.

Figure \ref{fig:tiers-relations} depicts how the Tier-1 API has been provided mainly as a Facade Pattern, facilitating the use and readability of the Tier-2 modules, reducing the signature of the higher-level API with the most common usage patterns. %
The Buffer is implemented as a Proxy Pattern to provide extra management features and a common interface for different type of containers, independently of the nature (C pointers, \Cpp{} containers) and its locality (host or device memory). 
Finally, the Strategy Pattern has been used in the pluggable scheduling system, where each scheduler is encapsulated as a strategy that can be easily interchangeable within the family of algorithms. Due to its common interface, new schedulers can be provided to the runtime. \newline

In summary, EngineCL is designed following an API and feature-driven development to achieve high external usability (API design) and internal adaptability to support new runtime features when the performance is not penalized. This is accomplished through a layered architecture and a set of core modules well profiled and encapsulated.

%% file: implementation.tex
\section{EngineCL Implementation}
\label{sec:Implementation}

\subsection{Multi-Threaded Architecture}

EngineCL has been developed in \Cpp{}, mostly using \Cpp{}11 modern features to reduce the overhead and code size introduced by providing a higher abstraction level.
Many modern features such as \textit{rvalue references}, \textit{initializer lists} or \textit{variadic templates} have been used to provide a better and simpler API while preserving efficient management operations internally.

When there is a trade-off between internal maintainability of the runtime and a performance penalty seen by profiling, it has been chosen an implementation with the minimal overhead in performance. As an example, when using \texttt{shared\_ptr} could induce unknown performance penalties internally, they have been discarded in favor of raw pointers. Also, other Design Patterns, such as the Chain of Responsibility, could be applied to the architecture to increase the maintainability, but they introduce higher overheads. Therefore, the pattern was not applied to preserve the maximum performance.

EngineCL has a multi-threaded architecture that combines the best measured techniques regarding OpenCL management of queues, devices and buffers. Some of the decisions involve atomic queues, parallel operations, custom buffer implementations, reusability of costly OpenCL functions, efficient asynchronous enqueueing of operations based on callbacks and event chaining. These mechanisms are used internally by the runtime and hidden from the programmer to achieve efficient executions and transparent management of devices and data.

\subsection{Optimizations}

The implementation follows feature-driven development to allow incremental features based on requested needs when integrating new vendors, devices, type of devices and benchmarks. Implementation techniques are profiled with a variety of OpenCL drivers from the major vendors (AMD, Intel, Nvidia and ARM) and versions (1.0, 1.1, 1.2, 2.0), but also in devices of different nature, such as integrated and discrete GPUs, CPUs, accelerators such as Xeon Phi and even FPGAs  (complex architecture that has been incorporated in EngineCL \cite{Guzman:2019}).

As examples of optimizations, two implementations have been incorporated to reduce overheads produced both in the initialization and closing stages of the program, mainly due to the use of OpenCL drivers in the analyzed infrastructures. These modifications are tagged as \textit{initialization} and \textit{buffer optimizations}.

The first optimization focuses on taking advantage of the discovery, listing and initialization of platforms and devices by the same thread (Runtime). In parallel, both the thread in charge of load balancing (Scheduler) and the threads associated with devices (Device) take advantage of this time interval to start configuring and preparing their resources as part of the execution environment. These threads will wait only if they have finished their tasks independent of the OpenCL primitives, instantiated by the Runtime. The runtime takes advantage of the same discovery and initialization structures to configure the devices before delegating them to the Scheduler and Device threads, which will be able to continue with the following stages. These optimizations reduce the execution time affecting the beginning and end of the program, due to the increase of the parallel fraction of the program as well as the reuse of the structures in memory, liberating the redundant OpenCL primitives.

On the other hand, some modifications have been made when instantiating and using both input and output buffers (Buffer). The variety of architectures as well as the importance of sharing memory strategies save costs when doing transfers and unnecessary complete bulk copies of memory regions, usually between main memory and device memory, but also between reserved parts of the same main memory (CPU - integrated GPU). By tweaking OpenCL buffer flags that set the direction and use of the memory block with respect to the device and program, OpenCL drivers are able, if possible, to apply underlying optimizations to the memory management.

\begin{figure*}[!h]
\centering
    \includegraphics[width=\textwidth]{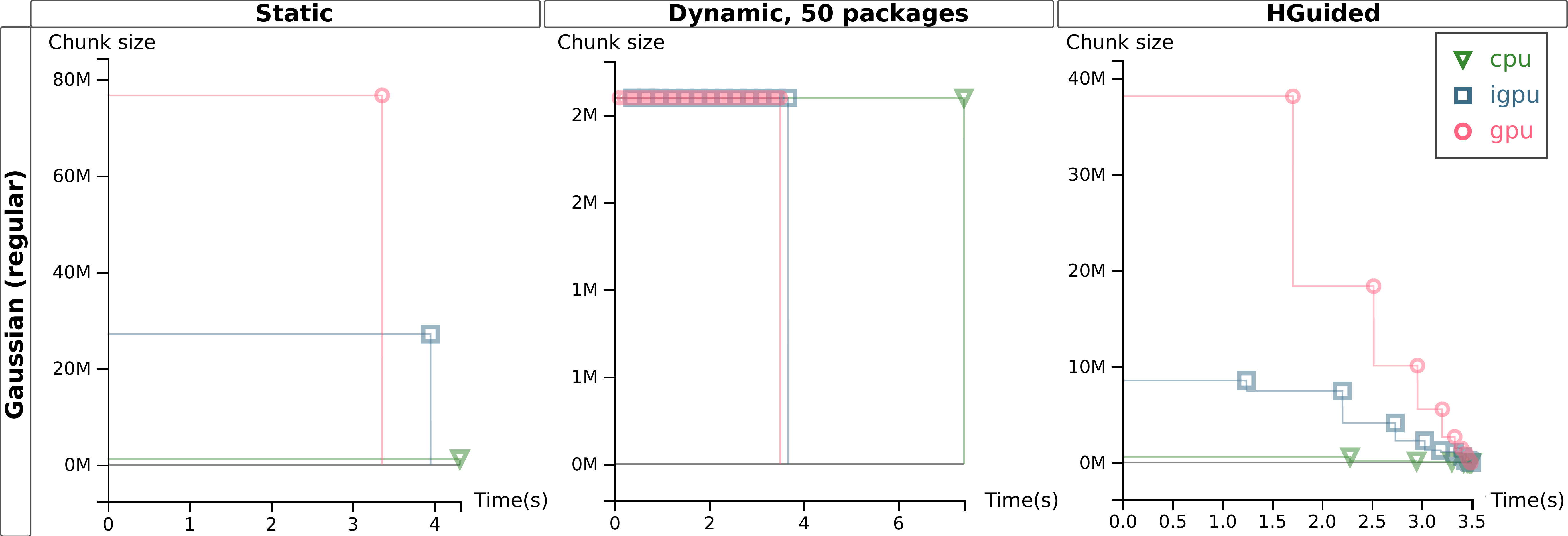}
\caption{Package distribution for every Load Balancing algorithm, using the EngineCL's Introspector module to extract the data from the execution. Every marker represents when a device computes a package in Gaussian (regular problem).}
\label{fig:lbaChunks}
\vspace{-3mm}
\end{figure*}

\begin{figure*}[!h]
\centering
    \includegraphics[width=\textwidth]{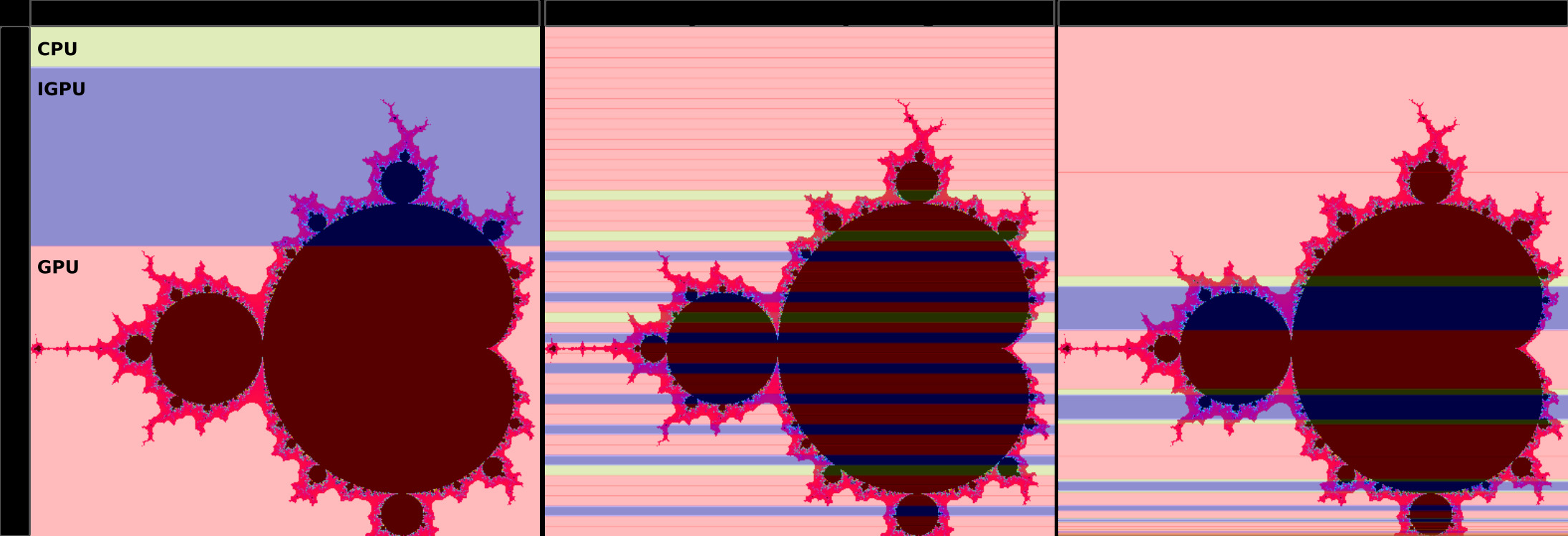}
\caption{EngineCL's Instrospector visual representation of the package distribution when computing Mandelbrot (irregular problem). The execution is done from top to bottom in each image. To help understanding, colored horizontal sections overlap the mandelbrot fractal (real computation), representing the chunk sizes computed by each device.}
\label{fig:lbaChunksIrregular}
\vspace{-3mm}
\end{figure*}


\subsection{Schedulers}
\label{sec:schedulers}

The EngineCL architecture allows to easily incorporate a set of schedulers, as it is shown in Figure \ref{fig:tiers-relations}. In this paper, three well-known schedulers are implemented in EngineCL \cite{Maat:2009, Guzman:2019, Nozal:2018}. The programmer can select one scheduler per kernel execution, depending on the characteristics and knowledge he has of the problem, data communication and architecture. Figure \ref{fig:lbaChunks} and \ref{fig:lbaChunksIrregular} depict the three load balancing algorithms in terms of chunk computation (chunk size and time) and visual representation of the computation (portion of the problem size computed per device), for Gaussian (regular) and Mandelbrot (irregular problem), respectively.
The algorithms used in this article are briefly described:

\paragraph{Static} This algorithm works before the kernel is executed by dividing the dataset in as many packages as devices are in the system. The division relies on knowing the percentage of workload assigned to each device, in advance. Then the execution time of each device can be equalized by proportionally dividing the dataset among the devices. It minimizes the number of synchronization points; therefore, it performs well when facing regular loads with  known computing powers that are stable throughout the dataset. However, it is not adaptable, so its performance might not be as good with irregular loads. This can be seen in the bottom left image: if the CPU receives the last part of the image, it will be highly imbalanced.

\paragraph{Dynamic} It divides the dataset in a given number of equal-sized packages. The number of packages is well above the number of devices in the heterogeneous system. During the execution of the kernel, a master thread in the host
assigning packages to the different devices, including the CPU. This algorithm adapts to the irregular behavior of some applications, like it is depicted in bottom center Mandelbrot computation. However, each completed package represents a synchronization point between the device and the host, where data is exchanged, and a new package is launched. This overhead has a noticeable impact on performance if the number of packages is high.

\paragraph{HGuided}

HGuided offers a variation over the Dynamic algorithm by establishing how the dataset is divided. The algorithm makes larger packages at the beginning and reduces the size of the subsequent ones as the execution progresses, as it is shown in the right side of both figures. Thus, the number of synchronization points and the corresponding overhead is reduced, while retaining a small package granularity towards the end of the execution to allow all devices to finish simultaneously.

Since it is an algorithm for heterogeneous systems the size of the packets is also dependent on the computing power of the devices. The size of the package for device $i$ is calculated as follows:

\begin{equation*}
  packet\_size_i = \floor[\bigg]{\frac{G_r\,P_i}{k\,n\,\sum_{j=1}^{n} P_j}}
\end{equation*}
where $k_i$ is an arbitrary constant. The smaller the $k$ constant, the faster decreases the packet size. Tweaking this constant prevents too large packet sizes when there are only a few devices, unbalancing the load.
$G_r$ is the number of pending work-groups and is updated with every package launch. The parameters of the HGuided are the computing powers and the minimum package size of the devices to be used. 
The minimum package size is the lower bound of the $packet\_size_i$, and it is dependent on the computing power of the devices, giving bigger package sizes in the most powerful devices.

%% file: api.tex
\section{API Utilization}
\label{sec:API}

This section describes two use cases of the EngineCL API. As the Section \ref{sec:Design} describes, the runtime has been thought from the beginning to provide a straightforward and flexible API from the point of view of the programmer. 
Both examples are real use cases, but they have been modified intentionally to show different API calls for demonstration purposes. As an example, the programmer will usually prefer a single call to \texttt{work\_items} than two consecutive calls to \texttt{global\_work\_items} and \texttt{local\_work\_items}.

The programmer starts by initializing the \texttt{EngineCL} and \texttt{Program}. The engine is the main element of the system because it manages devices, the application domain and extended features such as schedulers and introspection data (statistics of the execution). The engine handles well-known OpenCL concepts, such as the number of global and local work items.

The engine allows setting the devices to be used by masks (CPUs, GPUs, Accelerators, All devices in the system, any mixed combination, etc) or explicitly setting the platform and device. The latter mode is commonly used not only under development but also in production systems with many driver implementations (Pocl, Beignet, vendor specific, etc.) and when the programmer needs custom sets of devices or kernel specializations.

The concept of Program is decoupled from the runtime to help the programmer to understand it as an independent entity to be modified and to be easily extended to support multi-kernel executions. Therefore, it will allow establishing new parameters such as the concurrency of execution (many kernels at the same time) or linked buffers between programs (shared).

The API can be extended to support new features or to expose Tier-3 functionality to the above tiers, being able to use them directly without the need to access the EngineCL internal code. Finally, the API\footnote{API documentation available online at the URL address: \url{https://github.com/EngineCL/EngineCL}} is evolving as EngineCL integrates or supports new problems, data types, OpenCL features or devices, such as FPGAs, but the current examples show the core of its expressiveness and functionality.


\subsection{Case 1: Using only one device}

Listing \ref{list:APIUtilisationBinomial} shows how EngineCL is used to compute the benchmark Binomial Options with only a single device, the CPU. This example shows the explicit versions of some calls, such as \texttt{global} and \texttt{local} work items and a mixture of positional and aggregate kernel arguments. This is usually the first step to port OpenCL programs to EngineCL.

It starts reading the kernel, defining variables, containers (\Cpp{} \textit{vectors}) and OpenCL values like \textit{local} and \textit{global work size} (\texttt{lws}, \texttt{gws}). Then, the program is initialized based on the benchmark (\texttt{init\_setup}), in line 9 (\textit{L9}). The rest of the program is where EngineCL is instantiated, used and released.

The \texttt{engine} \textit{uses} the first CPU in the system by using a \texttt{DeviceMask} (\textit{L12}).
Then, the \textit{gws} and \textit{lws} are given by explicit methods (\textit{L14,15}).
The application domain starts by creating the \texttt{program} and setting the input and output containers with methods \texttt{in} and \texttt{out} (\textit{L17-19}).
With these statements the runtime manages and synchronizes the input and output data before and after the computation. 
The \texttt{out\_pattern} is set because the implementation of the Binomial OpenCL kernel uses a writing pattern of $\frac{1\ out\ index}{255\ work-items}$ (\textit{L21}), that is, 255 work-items compute a single out index. Then, the kernel is configured by setting its source code string, name and arguments.
Assignments are highly flexible, supporting aggregate and positional forms, and above all, it is possible to transparently use the variables and native containers (\textit{L23-29}).
The enumerated \texttt{LocalAlloc} is used to determine that the value represents the bytes of local memory that will be reserved, reducing the complexity of the API.
Finally, the runtime consumes the program and all the computation is performed (\textit{L32,34}). When the \texttt{run} method finishes, the output values are in the containers. 
 Optionally, errors can be checked and processed easily.


\begin{listing}[!htb]
\begin{snippet}[fontsize=\footnotesize]{c++}
auto kernel = file_read("binomial.cl");
auto samples = 16777216; auto steps = 254;
auto steps1 = steps + 1; auto lws = steps1;
auto samplesBy4 = samples / 4;
auto gws = lws * samplesBy4;
vector<cl_float4> in(samplesBy4);
vector<cl_float4> out(samplesBy4);

binomial_init_setup(samplesBy4, in, out);

ecl::EngineCL engine;
engine.use(ecl::DeviceMask::CPU); // 1 Chip

engine.global_work_items(gws);
engine.local_work_items(lws);

ecl::Program program;
program.in(in);
program.out(out);

program.out_pattern(1, lws);

program.kernel(kernel, "binomial_opts");
program.arg(0, steps); // positional by index
program.arg(in); // aggregate
program.arg(out);
program.arg(steps1 * sizeof(cl_float4),
            ecl::Arg::LocalAlloc);
program.arg(4, steps * sizeof(cl_float4),
            ecl::Arg::LocalAlloc);

engine.use(std::move(program));

engine.run();

// if (engine.has_errors()) // [Optional lines]
//   for (auto& err : engine.get_errors())
//     show or process errors
\end{snippet}
\caption{EngineCL API used in Binomial benchmark.}
\label{list:APIUtilisationBinomial}
\end{listing}

\subsection{Case 2: Using several devices}

NBody program shows a more advanced example where EngineCL really excels. Listing \ref{list:APIUtilisationNBody} depicts EngineCL computing the NBody benchmark using three devices of the system: CPU, GPU and Xeon Phi. Because of that, the engine is configured to use one of the provided schedulers: the static approach. It will automatically balance the load based on the proportions given to the devices used. Moreover, it uses kernel specialization for different devices, getting the maximum performance per device, but also using the generic kernel for maximum compatibility.

Like in the previous use case, the benchmark is initialized up to line 13 (reading kernels, using \Cpp{} containers, etc.). Then, three kernels are used: a common version, a specific implementation for GPUs and a binary kernel built for the Xeon Phi (\textit{L1-3}). The \texttt{Device} class from the Tier-2 allows more features like platform and device selection by index (\texttt{platform, device}) and specialization of kernels and building options.
Three specific devices
are instantiated, two of them with special kernels (source and binary) by just giving to them the file contents (\textit{L17}). After setting the work-items in a single method, the runtime is configured to use the \texttt{Static} scheduler with different work distributions for the CPU, Phi and GPU (\textit{L23,24}).
Finally, the program is instantiated without any out pattern, because every work-item computes a single output value,
and the seven arguments are set in a single method, increasing the productivity even more (\textit{L33}).

\begin{listing}[!tb]
\begin{snippet}[fontsize=\footnotesize]{c++}
auto kernel = file_read("nbody.cl");
auto gpu_kernel = file_read("nbody.gpu.cl");
auto phi_kernel_bin =
  file_read_binary("nbody.phi.cl.bin");
auto bodies = 512000; auto del_t = 0.005f;
auto esp_sqr = 500.0f; auto lws = 64;
auto gws = bodies;
vector<cl_float4> in_pos(bodies);
vector<cl_float4> in_vel(bodies);
vector<cl_float4> out_pos(bodies);
vector<cl_float4> out_vel(bodies);

nbody_init_setup(bodies, del_t, esp_sqr, in_pos,
                 in_vel, out_pos, out_vel);

ecl::EngineCL engine;
engine.use(ecl::Device(0, 0),
           ecl::Device(0, 1, phi_kernel_bin),
           ecl::Device(1, 0, gpu_kernel));

engine.work_items(gws, lws);

auto props = { 0.08, 0.3 };
engine.scheduler(ecl::Scheduler::Static(props));

ecl::Program program;
program.in(in_pos);
program.in(in_vel);
program.out(out_pos);
program.out(out_vel);

program.kernel(kernel, "nbody");
program.args(in_pos, in_vel, bodies, del_t,
             esp_sqr, out_pos, out_vel);

engine.program(std::move(program));

engine.run();
\end{snippet}
\caption{EngineCL API used in NBody benchmark.}
\label{list:APIUtilisationNBody}
\end{listing}

As it is shown, EngineCL manages both programs with an easy and similar API, but completely changes the way it behaves: Binomial is executed completely in the CPU, while NBody is computed using the CPU, Xeon Phi and GPU with different kernel specializations and workloads. Platform and device discovery, data management, compilation and specialization, synchronization and computation are performed transparently for the programmer in a few lines. As it was depicted in Section \ref{sec:Design} in Figure \ref{fig:overview} and later exposed in Section \ref{sec:Validation}, EngineCL saves hundreds to thousands of lines of code to manage (efficiently) all the operations here exposed to compute each program, but even more when using all the available resources of the heterogeneous system. EngineCL only needs a single line to incorporate a new device to the co-execution.

%% file: methodology.tex
\section{Methodology}
\label{sec:Methodology}

EngineCL has been validated both in terms of usability and performance.

\subsection{System Setup}
The experiments have been carried out using two different machines to validate both code portability and performance of EngineCL.

\paragraph{Batel} is a heterogeneous system composed of two Intel Xeon E5-2620 CPUs with six cores that can run two threads each at 2.0 GHz and 16 GBs of DDR3 memory. The CPUs are connected via QPI, which allows OpenCL to detect them as a single device. Therefore, throughout the remainder of this document, any reference to the CPU includes both processors. Moreover, it has one NVIDIA Kepler K20m GPU with 13 SIMD lanes (or SMs in NVIDIA terminology) and 5 GBytes of VRAM. And an Intel Xeon Phi KNC 7120P, with 61 cores and 244 threads. These are connected to the system using independent PCI 2.0 slots.

\paragraph{Remo} is a machine composed of an AMD A10-7850K APU and Nvidia GeForce GTX 950 GPU. The CPU has 2 cores and 2 threads per core at 3142 Mhz with only two cache levels, exposing 4 OpenCL compute units. The APU's on-chip GPU is a GCN 2.0 Kaveri R7 DDR3 with 512 cores at 720 Mhz with 8 compute units. Finally, the Nvidia discrete GPU has 768 cores at 1240 Mhz with GDDR5, providing 6 compute units.

It is interesting to emphasize that with these two nodes it is possible to test the versatility of EngineCL for 6 different types of devices: Intel CPU, AMD CPU, commodity GPU, HPC GPU, integrated GPU and Intel Xeon Phi. EngineCL has also been tested with FPGAs, with good results that can be seen in \cite{Guzman:2019}.

\subsection{Benchmarks}
 Five benchmarks have been used to show a variety of scenarios regarding the ease of use, overheads compared with a native version in OpenCL \Cpp{} and performance gains when multiple heterogeneous devices are co-executed. Table \ref{tbl:properties} shows the properties of every benchmark. Gaussian, Binomial, Mandelbrot and NBody are part of the AMD APP SDK, while Ray is an open source Raytracer implementation\cite{EngineCLBenchsuite}. Three different raytracing scenes (lights and objects) with different complexities are provided to be benchmarked when load balancing.


These five benchmarks are selected because they provide enough variety in terms of OpenCL development issues, regarding many parameter types, local and global memory usage, custom structs and types, number of buffers and arguments, different local work sizes and output patterns. The amount of properties, computing patterns and use cases are relevant because they provide enough diversity to compare EngineCL with OpenCL both in terms of overheads and usability. Moreover, the worst-case scenario is applied to EngineCL: a single device is used. 
All benchmarks are fair, executing the same kernels and using the OpenCL primitives efficiently. When load balancing, every device executes the same kernel, far from the best possible co-execution result, but strict with the bench-suite.






\subsection{Metrics}
The validation of usability is performed with eight metrics based on a set of studies (\cite{APIVisual:2009}, \cite{MaintenanceMetrics:2003}, \cite{MeasuresAPI:2013}, \cite{APIUsability:2015}).
These metrics determine the usability of a system and the programmer productivity, because the more complex the API is, the harder it is to use and maintain the program.

The \textit{McCabe's cyclomatic complexity} (\textit{CC}) measures the number of linearly independent paths. It is the only metric that is better the closer it gets to 1, whereas for the rest a greater value supposes a greater complexity. The \textit{number of \Cpp{} tokens} (\textit{TOK}) and \textit{lines of code} (\textit{LOC}, via \textit{tokei}) determine the amount of code. The \textit{Operation Argument Complexity} (\textit{OAC}) gives a summation of the complexity of all the parameters types of a method, while \textit{Interface Size} (\textit{IS}) measures the complexity of a method based on a combination of the types and number of parameters. The maintainability worsens the more parameters and more complex data types are manipulated. On the other side, \textit{INST} and \textit{MET} measure the number of \textit{Structs/Classes} instantiated and methods used, respectively. Finally, the \textit{error control sections} (\textit{ERRC}) measures the amount of sections involved with error checking.

A ratio of $\frac{OpenCL}{EngineCL}$ is calculated to show the impact in usability per benchmark and metric, except for \textit{CC} because it is a qualitative metric with zero as best value.

\begin{table}
\begin{center}
\caption{Benchmarks and variety of properties used in the validation.}
\label{tbl:properties}
\scriptsize
\renewcommand{\arraystretch}{1.5}
\begin{tabular}{@{\hskip 1mm}l@{\hskip 2mm}@{\hskip 2mm}>{\columncolor[gray]{0.95}}c@{\hskip 2mm}@{\hskip 2mm}c@{\hskip 2mm}@{\hskip 2mm}>{\columncolor[gray]{0.95}}c@{\hskip 2mm}@{\hskip 2mm}c@{\hskip 2mm}@{\hskip 2mm}>{\columncolor[gray]{0.95}}c@{\hskip 2mm}}
  \parbox[c][1.7cm]{2.2cm}{\vspace{1.2cm} \textbf{Property}}
  & \rotatebox[origin=s]{65}{\centering \textbf{Gaussian}}
  & \rotatebox[origin=c]{65}{\parbox[s]{13mm}{\centering \textbf{Ray}}}
  & \rotatebox[origin=c]{65}{\centering \textbf{Binomial}}
  & \parbox[c][1.8cm]{0.7cm}{\rotatebox[origin=c]{65}{\centering \textbf{Mandelbrot}\vspace{3mm}}}
  & \rotatebox[origin=c]{65}{\parbox[c]{13mm}{\centering \textbf{NBody}}} \\ \hline
Local Work Size & 128 & 128 & 255 & 256 & 64\\ \hline
Read:Write buffers & 2:1 & 1:1 & 1:1 & 0:1 & 2:2\\ \hline
Out pattern & 1:1 & 1:1 & 1:255 & 4:1 & 1:1\\ \hline
Number of kernel args & 6 & 11 & 5 & 8 & 7\\ \hline
Use local memory & no & yes & yes & no & no\\ \hline
Use custom types & no & yes & no & no & no\\ \hline
\end{tabular}
\end{center}
\vspace{-7mm}
\end{table}

Regarding the performance evaluation two types of experiments are presented. The first measures the overhead of EngineCL compared with OpenCL \Cpp{} when using a single device, increasing the problem size. Due to the small overhead between EngineCL and OpenCL, a single background process can interfere with the results. Therefore, the experiments have been carried out removing every non-necessary process of the system (journal and periodic tasks), establishing user-defined CPU governors (fixed frequencies) and increasing the batch of executions to reduce the noise of the system.
The analysis focuses on small problem sizes, because it is where the relevant overheads appear. The minimum problem sizes are selected based on the computing power of every device, being reasonable for each benchmark
and usually around the second of execution, including initialization, management and releasing.
Then, the size increases per device and benchmark until the overheads are stabilized or when the execution time is prohibitive, such as CPU reaching more than 100 seconds of execution or GPU being memory-bounded. As a result, it shows the overall trend.

The time overhead, expressed as percentage, is computed as the ratio between the difference of the response times in the execution of the same kernel for both EngineCL ($T_{ECL}$) and the native version ($T_{OCL}$), as follows: $Overhead = \frac{T_{ECL} - T_{OCL}}{T_{OCL}} \cdot 100$.

The second analyzes the co-execution performance when using different scheduling configurations in a heterogeneous system composed of three different devices. Each program uses a single problem size, given by the completion time of around 10 seconds in the fastest device (GPU) for Batel, and 7 seconds in the fastest device (GPU) for Remo.




To evaluate the performance of EngineCL the total response time and the time per device are measured, including the device initialization and management, input data and results communications. Then three metrics are computed: balance, speedup and efficiency.

To measure the effectiveness of load balancing, we calculate the balance as $ \frac{ {T_{FD}} }{ {T_{LD}} }$, where $T_{FD}$  and $T_{LD}$ are the execution time of the device that finish at first and last, respectively. Thus, it is $1$ if all finish at the same time.

For the latter two metrics, the baseline is always the fastest device running a single invocation of the kernel (GPU in both nodes).
Due to the heterogeneity of the system and the different behavior of the benchmarks, the maximum achievable speedups depend on each program. These values derive from the response time $T_i$ of each device:

\vspace{-2mm}
\begin{equation*}
   S_{max} =  \frac{1}{max_{i=1}^n\{T_i\}} \sum_{i=1}^n T_i
   \label{smax}
\end{equation*}

Additionally, the efficiency of the heterogeneous system has been computed as the ratio between the maximum achievable speedup and the empirically obtained speedup for each benchmark. $Eff = \frac{S_{real}}{S_{max}}$.

The scheduling configurations are grouped by algorithm. The first two bars represent the Static algorithm varying the order of delivering the packages to the devices. The one labeled \emph{Static} delivers the first chunk to the CPU, the second to the iGPU/PHI (depending on the node) and the last one to the GPU, while in the \emph{Static rev} the order is \textit{GPU-iGPU/PHI-CPU}. The next two show the Dynamic scheduler configured to run with 50 and 150 chunks. Finally, the latter presents the HGuided algorithm.

To guarantee integrity of the results when doing the load balancing experiments, 60 executions are performed per case, divided in 3 sets of no consecutive executions. Every set of executions performs 20 iterations contiguously without a wait period, discarding an initial execution to avoid warm-up penalties in some OpenCL drivers and devices. When measuring the overheads, the experiments are modified to 300 executions, 2 sets and 100 iterations.

%% file: validation.tex

\section{Validation}
\label{sec:Validation}

\input{validation_usability_table_chart.tex}

We perform the necessary experimentation to try to answer four questions. First, how easy and maintainable it is to program heterogeneous systems with EngineCL compared to OpenCL, thanks to the use of a set of metrics that are the state of the art in research in Software Engineering. Second, considering one of the worst scenarios for EngineCL, how much overhead EngineCL has over OpenCL when only one device is used to offload the computation. Third, considering the co-execution of all devices in a system, how good EngineCL is balancing the load between them. And finally, taking into account the EngineCL runtime, its design decisions and load balancing algorithms, how much performance and efficiency is obtained when all the devices in the heterogeneous system are fully exploited.

\subsection{Usability}
\label{sec:Usability}

This section shows the experiments performed to evaluate the usability introduced by EngineCL when a single device is used. Table \ref{tbl:usability} presents the values obtained for every benchmark (rows) in every of the eight metrics (columns). Also, the typical (average) ratio per metric for the set of programs is presented as a chart.

For every program, the maintainability and testing effort is reduced drastically, as can be seen
with \textit{CC}, reaching the ideal cyclomatic complexity, or \textit{ERRC}. The savings in error checking are on average 21 times less by using EngineCL, reducing the visual complexity of alternate paths for error control. 




\begin{figure*}[!h]
   \centering
   \includegraphics[width=0.99\textwidth]{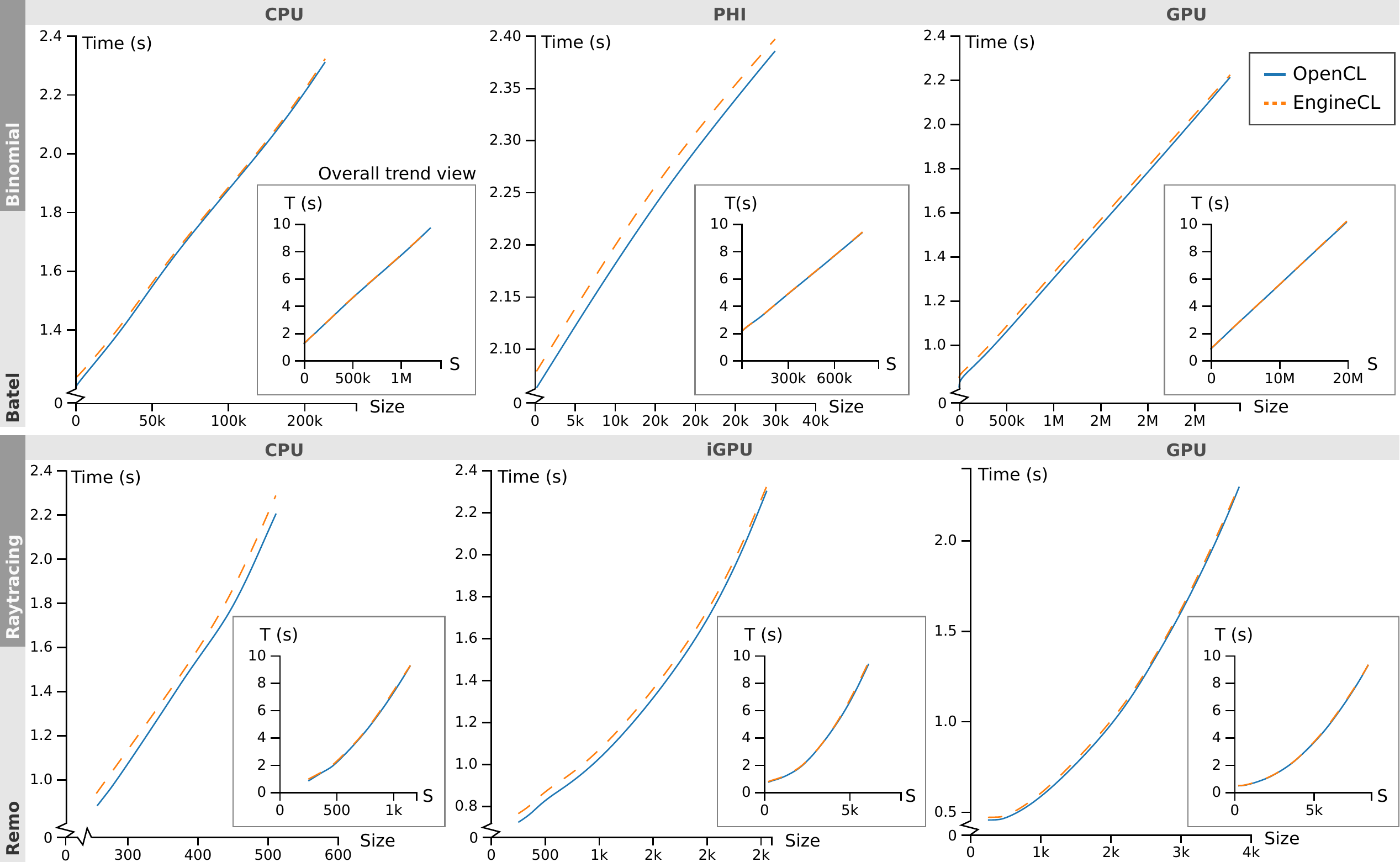}
   \caption{Overheads of EngineCL compared with OpenCL for each device in the system. Worst overhead results are found when computing Binomial Options in Batel for the CPU (top) and Raytracing in Remo for the CPU and GPU (bottom).}
   \label{fig:overheadsBinomialRay}
   \vspace{-3mm}
\end{figure*}

\begin{figure*}[!h]
   \centering
   \includegraphics[width=0.99\textwidth]{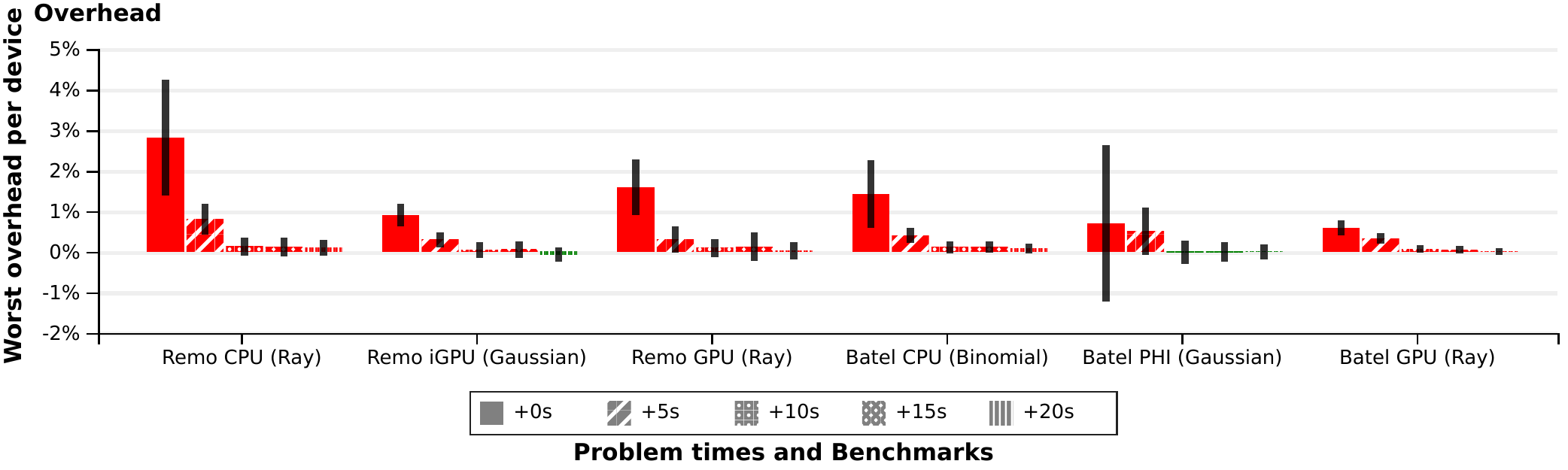}
   \caption{Worst overheads per device. The overheads decrease with longer execution times. As it is said in Section \ref{sec:Methodology}, every device and benchmark has its minimum problem size. Every bar represents the overheads for specific execution times added to the execution time for each minimum problem size (i.e. Remo CPU for Ray are 0.8, 5.8, 10.8, 15.8 and 20.8, while Batel CPU for Binomial are 1.2, 6.2, \ldots{} and 21.2).}
   \label{fig:overheadsWorstCases}
   \vspace{-2mm}
\end{figure*}

The code density and complexity of the operations are reduced between 7.3 to 8.5 times compared with OpenCL, as it is shown with
TOK, OAC and IS.
In programs like Ray the \textit{OAC} ratio is greater than in \textit{TOK}, because the number of parameters grows in both implementations, but managing complex types is harder in OpenCL.

The number of classes instantiated and used methods are around 5 and 2 times less than in the OpenCL implementation, mainly because it has been deliberately instantiated Tier-2 classes. 

As a summary, EngineCL has excellent results in maintainability, implying less development effort. Thanks to its API usability, the programmer is able to focus on the application domain, and its productivity is boosted by hiding complex decisions, operations and checks related with OpenCL. These excellent results take into account the worst-case for EngineCL: using a single device.


\subsection{Overhead of EngineCL}
\label{sec:Performance}

This section presents results of experiments performed to evaluate the overhead introduced by EngineCL when a single kernel is executed in a single device.

Figure \ref{fig:overheadsBinomialRay} shows the execution times measured for both OpenCL and EngineCL completion time for different problem sizes. Only the worst EngineCL over OpenCL overheads per node are shown, one is regular (Binomial) and other is irregular (Ray). The maximum overhead value is produced with the CPU in the Remo node, with very small problem sizes. The maximum overhead measured is 2.8\%, while the average value obtained for the minimum problem size for all benchmarks is 1.3\%. The overall trend view shows the problem scalability along with the overhead introduced by EngineCL, that it is minimized with larger problem sizes. The figure highlights the execution times for the smallest problem sizes, showing the slightly differences between runtimes.


\begin{figure*}[!h]
   \centering
   \includegraphics[width=0.49\textwidth]{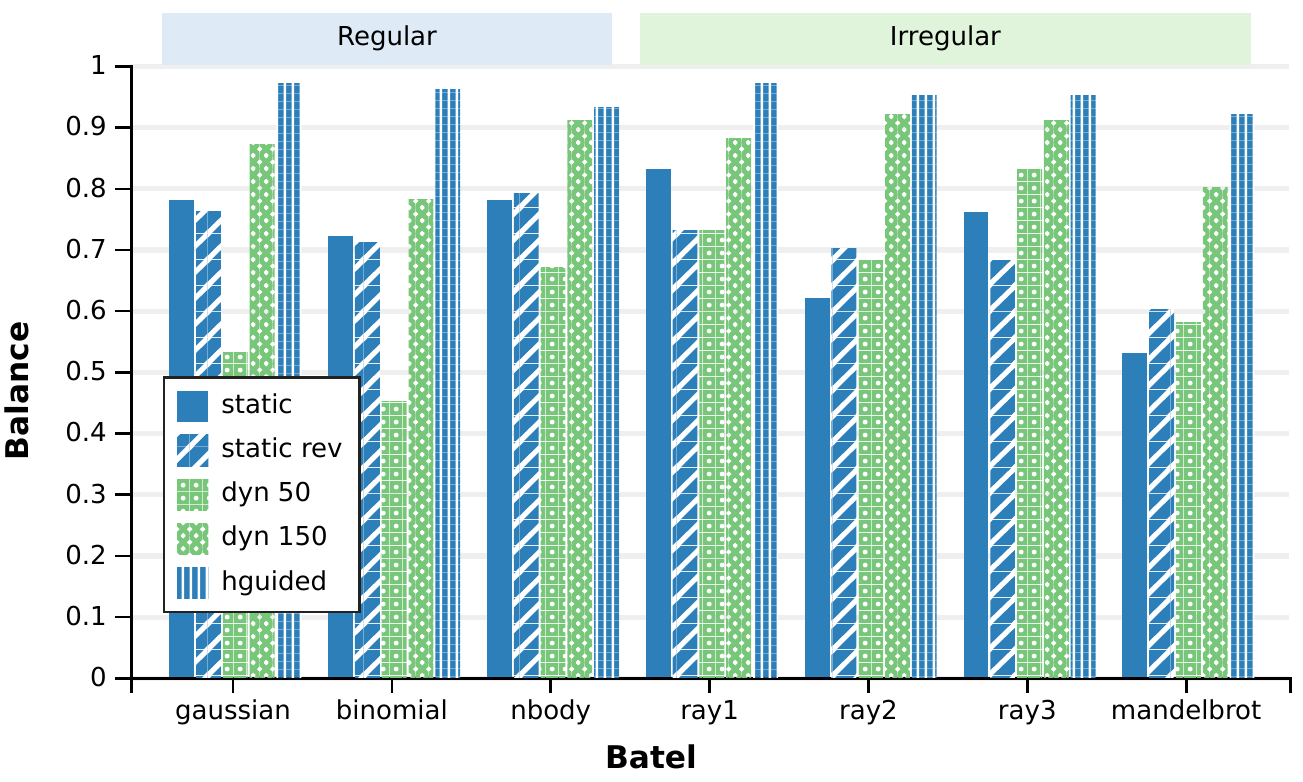}
   \includegraphics[width=0.49\textwidth]{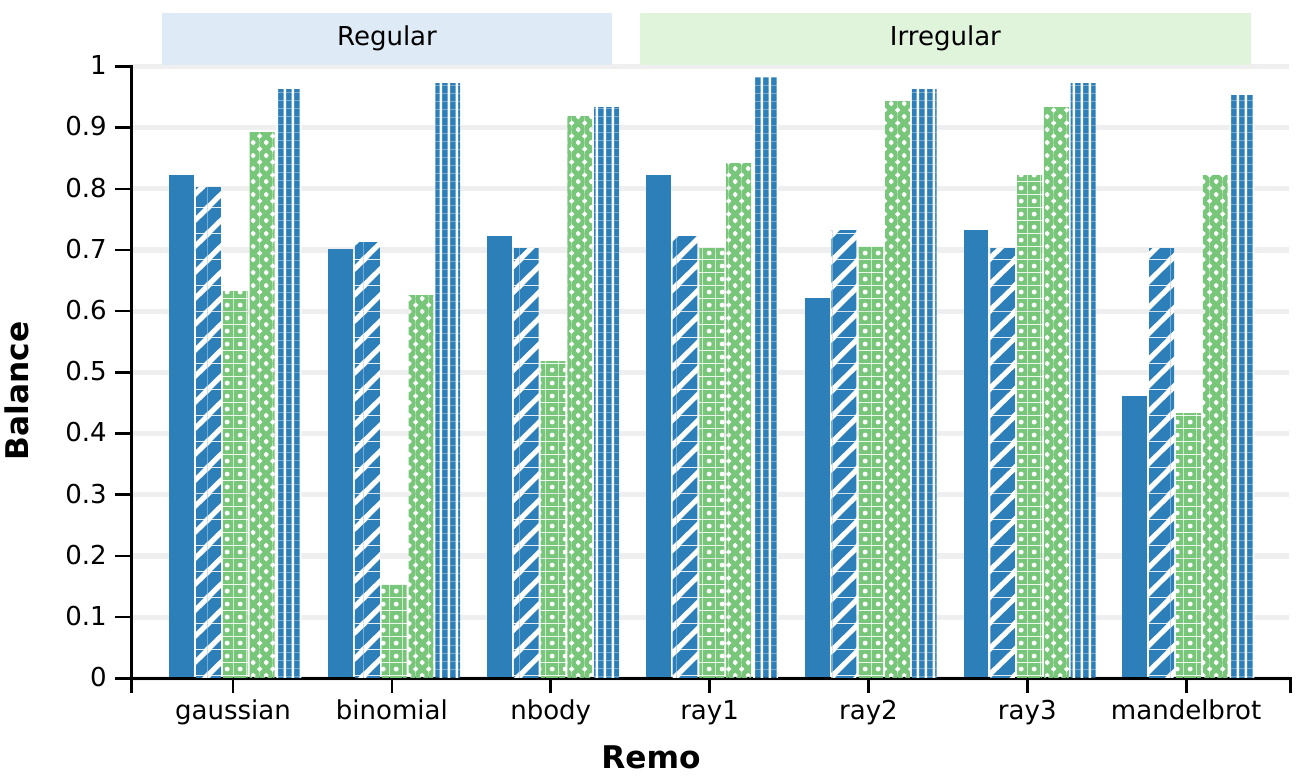}
   \vspace{-1mm}
   \caption{Balancing of the system per benchmark and scheduling configuration.}
   \label{fig:balance}
   \vspace{-2mm}
\end{figure*}


On the other side, Figure \ref{fig:overheadsWorstCases} depicts the worst overheads per device and benchmark, including the variability (errors, standard deviation). Analyzing each device separately, it can be observed that the worst results are obtained in the Remo CPU. This is reasonable since EngineCL also runs on the CPU, that has only 2 cores and 4 threads. Therefore, its multi-threaded architecture interferes with the execution of benchmarks, stealing them computing capacity. This behavior is highly mitigated in the Batel CPU, where the threads used by the runtime does not interfere with the 24 computing threads of the CPU. Regarding the discrete devices, the differences between devices are mainly dependent on the driver implementation and how it is affected by the multi-threaded and optimized architecture of EngineCL. The commodity Remo GPU has the highest overhead between the discrete devices, up to 1.59\%, but quickly reducing it with larger problem sizes. There are cases, like the Xeon Phi, in which the driver and device produces high variability in the results, probably produced by the amount of host threads that are spawned (up to 24).

Two conclusions can be drawn from the results as a whole. On the one hand, the overhead introduced by the EngineCL runtime is 1.3\% on average for all evaluated devices. On the other hand, that EngineCL scales very well with the execution time, so that the overhead decreases significantly as the execution time of the application increases.

\subsection{Load Balancing}
The next question to be analyzed is whether EngineCL successfully distributes the workload among the devices of the heterogeneous system. To this end, Figure \ref{fig:balance} presents the \emph{Load Balance}, defined as the ratio of the response times of the first and last devices to conclude its work. The ideal value for this metric is one, meaning that all devices finished simultaneously and the maximum utilization of the machine was attained.

Based on these results, three general conclusions can be outlined. Firstly, EngineCL successfully balances the workload in the two systems analyzed. The mean value of the balance is 0.96, very close to 1.0, with maximum values of 0.98, for example in Gaussian (Batel) and Ray1 (Remo). Secondly, HGuided is the algorithm that offers the best results in all the scenarios studied, in both Batel and Remo, and for both regular and irregular applications. Finally, it can also be seen the great relevance of selecting a suitable load balancing algorithm, since otherwise very large imbalances can occur as shown in the cases of static algorithms in Mandelbrot or dynamic approaches with a few packages for Binomial.

Regarding the rest of the algorithms, it can be observed that both static algorithms have a very similar behavior in regular applications, as expected. Nevertheless, they present important differences in the irregular ones, for instance Mandelbrot in Remo. Besides, their behavior depends completely on each case, as can be seen in the cases of Ray1 (static is better) and Ray2, where the reverse gets better results. Finally, the dynamic algorithm always achieves the best-balanced results with the greatest number of packets. However, as it can be seen below, this does not always mean the best performance.

\begin{figure*}[!t]
   \centering
   \includegraphics[width=0.49\textwidth]{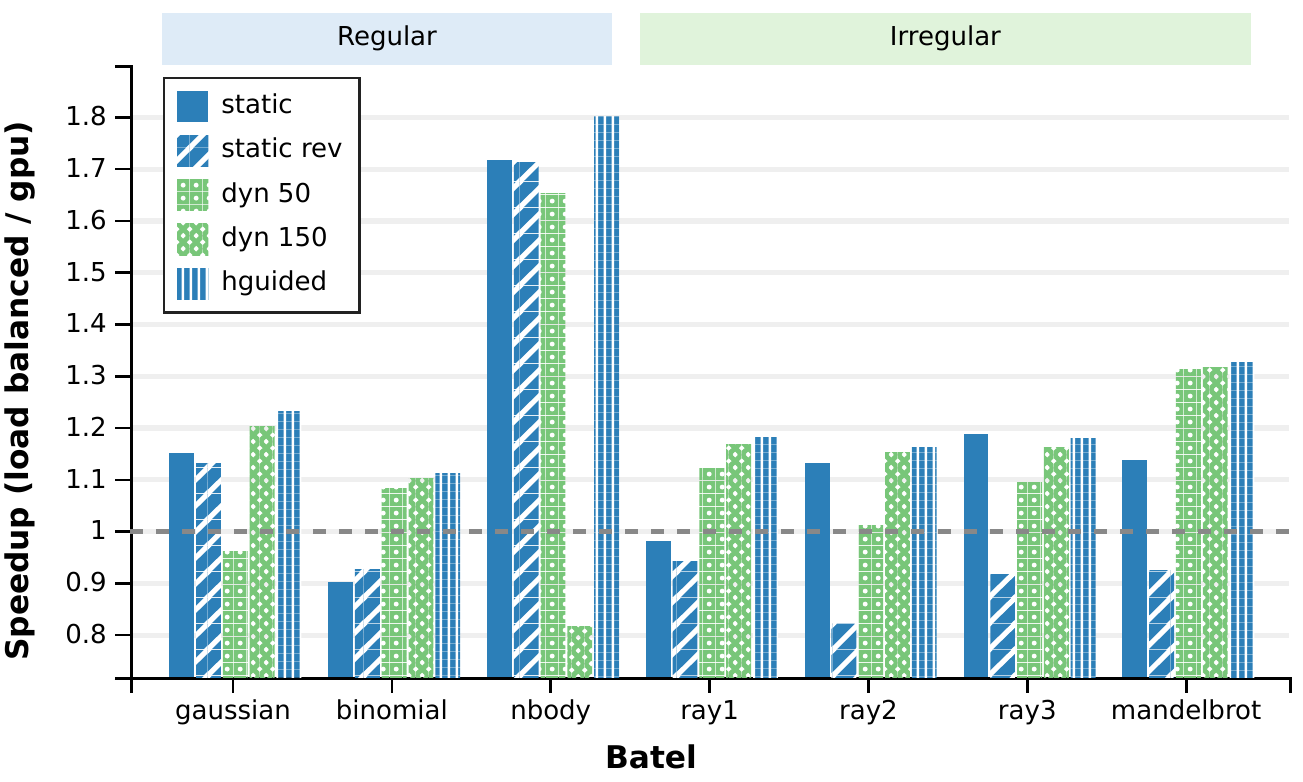}
   \centering
   \includegraphics[width=0.49\textwidth]{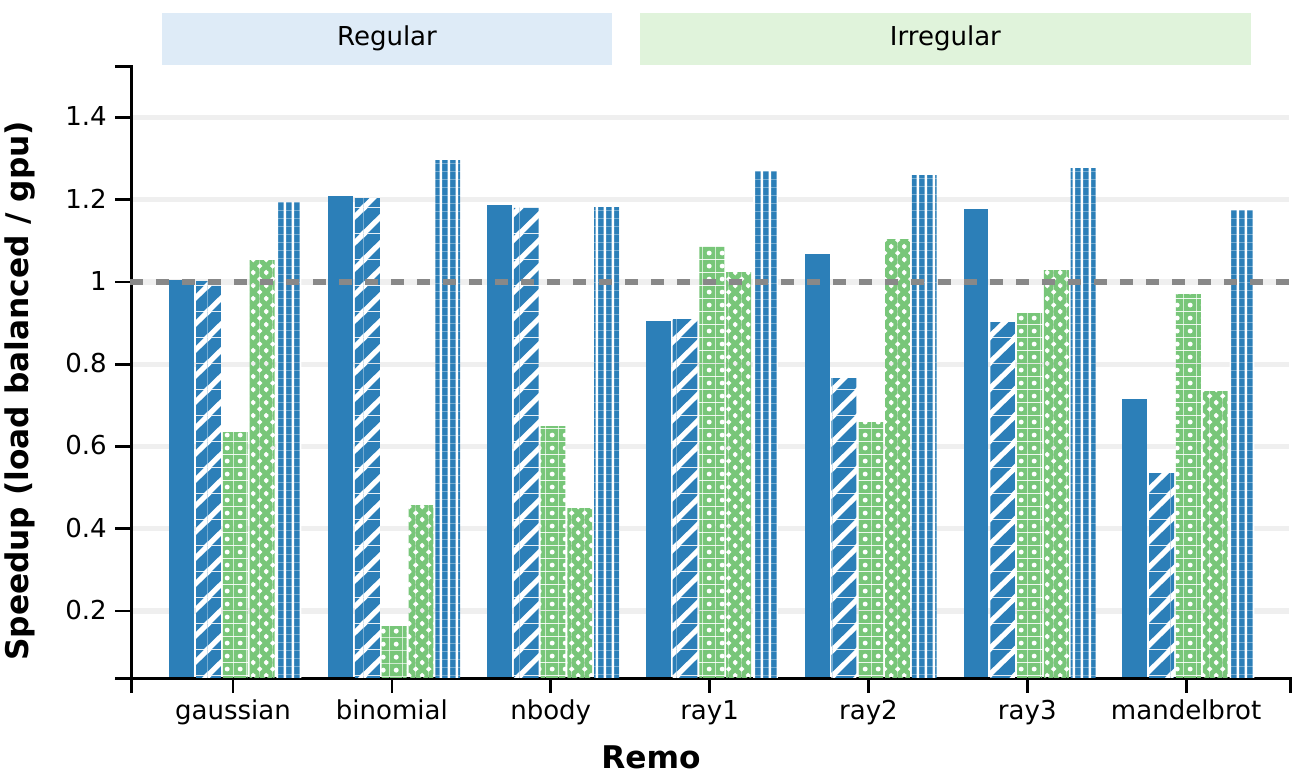}
   \vspace{-1mm}
   \caption{Speedups for every scheduler compared with a single GPU.}
   \label{fig:speedup}
   \vspace{-2mm}
\end{figure*}

\begin{figure*}[!t]
   \centering
   \includegraphics[width=0.49\textwidth]{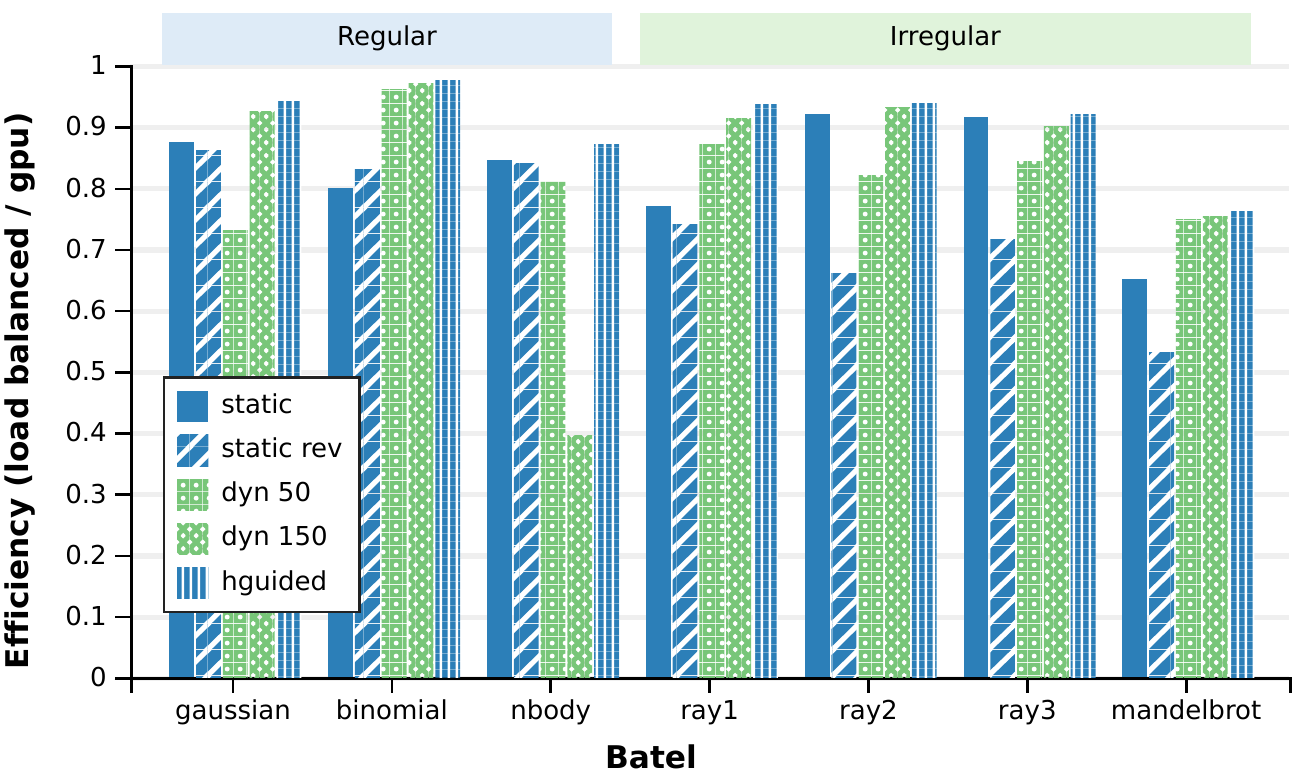}
   \centering
   \includegraphics[width=0.49\textwidth]{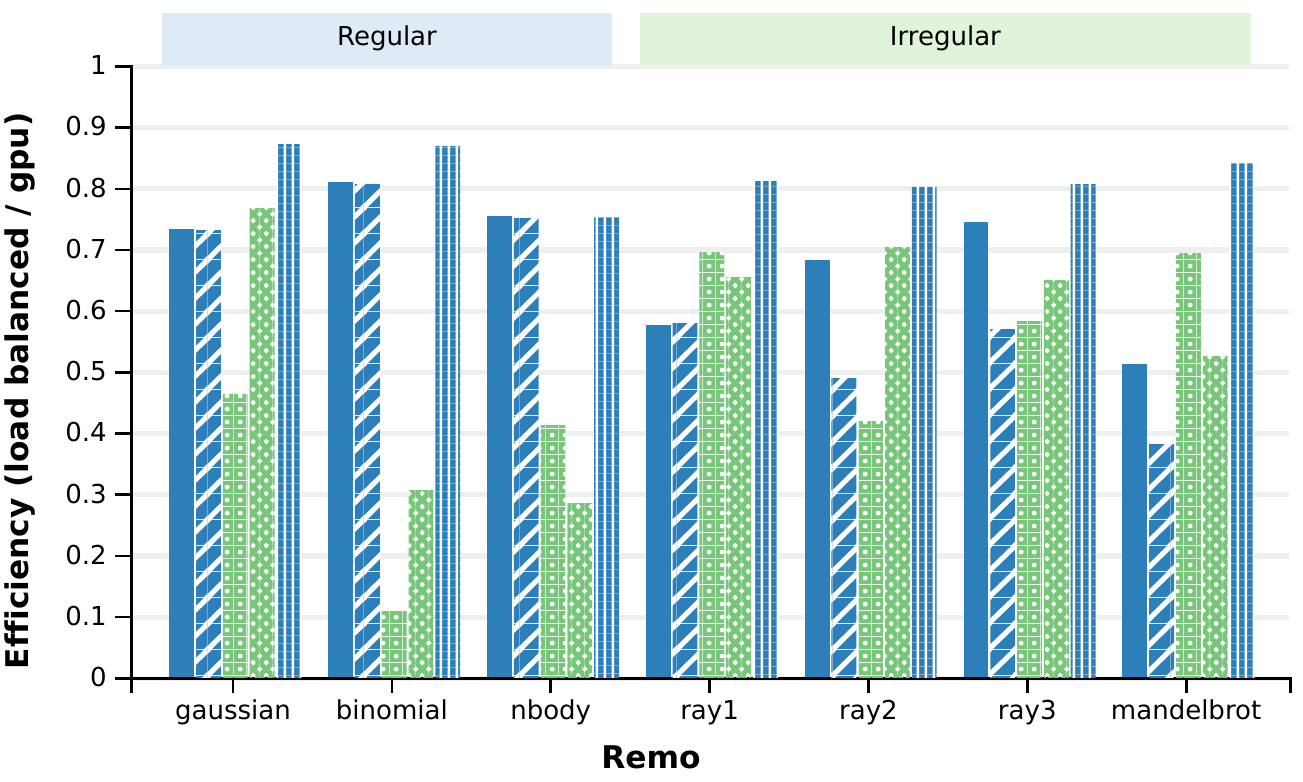}
   \vspace{-1mm}
   \caption{Efficiency of the system when adding the CPU and Xeon Phi.}
   \label{fig:efficiency}
   \vspace{-2mm}
\end{figure*}

\subsection{Performance}
The performance results achieved in the heterogeneous systems (Batel left and Remo right) with different load balancing algorithms are shown in Figure \ref{fig:speedup} and \ref{fig:efficiency}, where the speedups and efficiency are depicted, respectively.
The speedups are due to the co-execution of the benchmarks simultaneously on all the devices of the heterogeneous system, compared with only using the fastest device in each node, that is the GPU on both nodes. The efficiency gives an idea of how the system is utilized. A value of 1.0 represents that all the devices have been working all the time.

The main conclusion that can be drawn is that, for all benchmarks and both nodes, co-execution provides performance improvements over the baseline. The magnitude of the improvements will depend on the computing power of the devices of the system. On the other hand, efficiency figures show that EngineCL can exploit co-execution very efficiently. This is an excellent result, taking into account the great difference in computing power that exists between the devices of the nodes employed.

But to achieve these improvements, it is necessary to select an appropriate load balancing algorithm. As can be seen in the figures, HGuided achieves the best results for all the scenarios analyzed, with an average efficiency of 0.89 in Batel and 0.82 in Remo. Therefore, EngineCL can adapt to different kinds of loads and computing nodes, obtaining outstanding performance.

\begin{figure*}[!t]
   \centering
   \includegraphics[width=0.49\textwidth]{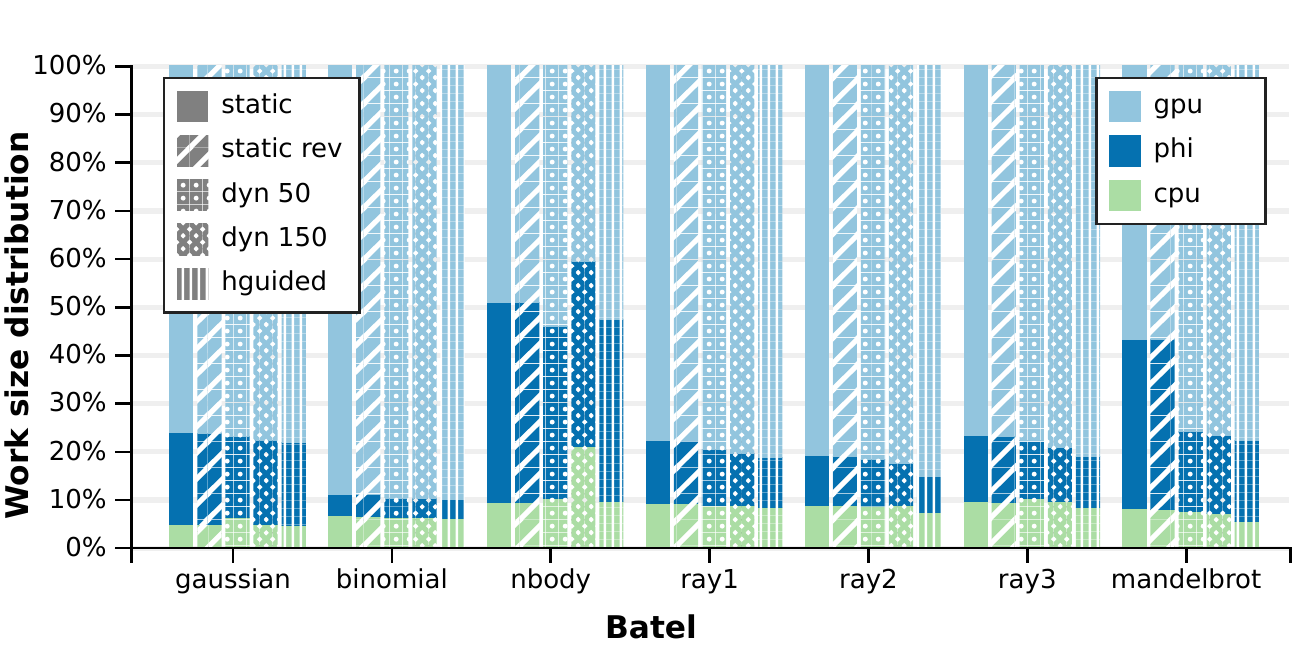}
   \includegraphics[width=0.49\textwidth]{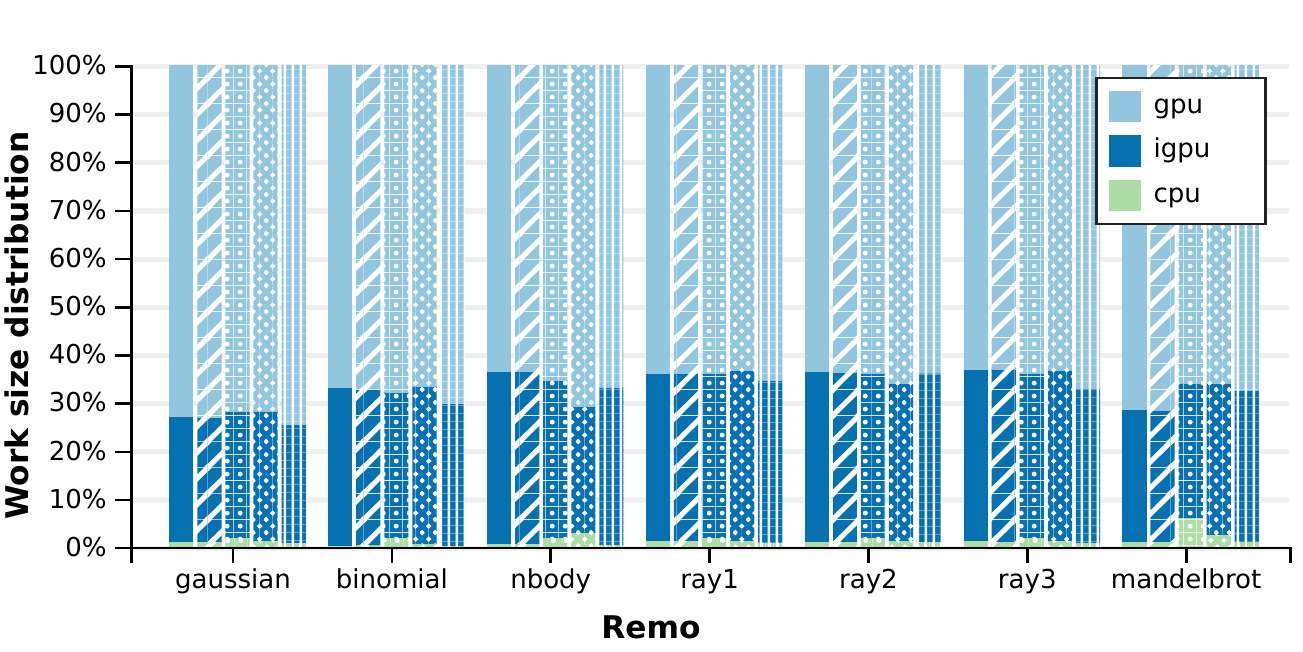}
   \vspace{-1mm}
   \caption{Work size distribution per device, benchmark and scheduler.}
   \label{fig:worksizes}
   \vspace{-2mm}
\end{figure*}

\begin{figure*}[!t]
   \centering
   \includegraphics[width=0.49\textwidth]{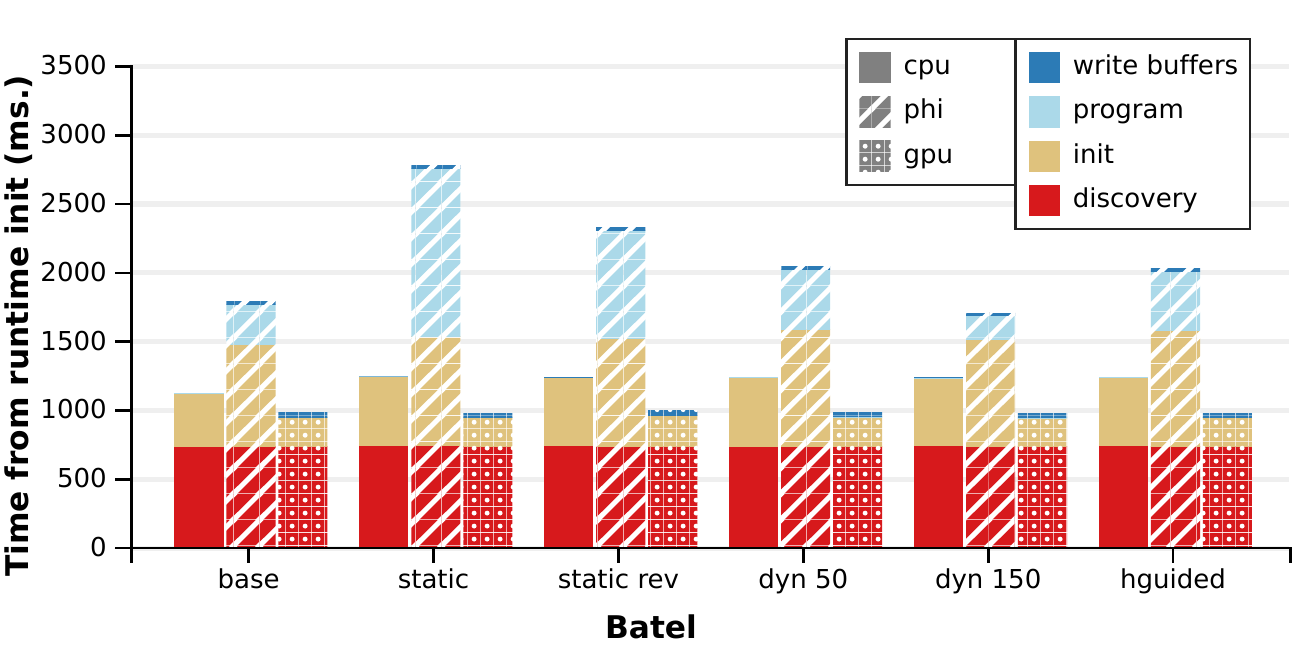}
   \centering
   \includegraphics[width=0.49\textwidth]{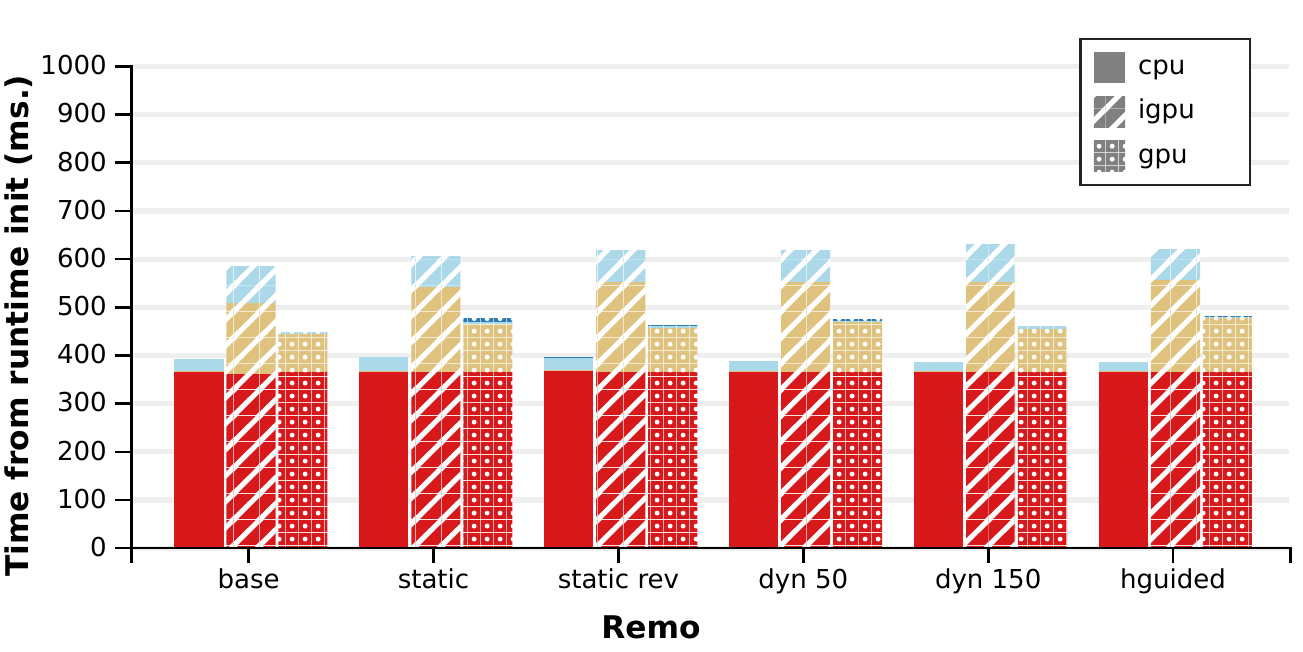}
   \vspace{-1mm}
   \caption{Binomial timings before the computation phase.}
   \label{fig:timingsBinomial}
   \vspace{-2mm}
\end{figure*}

Analyzing the speedups and efficiencies in detail, Static delivers good results in regular applications, with consistent efficiencies between 0.73 and 0.87, regardless of the order of the devices. Binomial in Batel is an exception that will be explained later, due to the Xeon Phi. However, in irregular applications the results are much more erratic, because it does not adapt to these irregularities, such as Ray1 (0.76) and Ray2 (0.92). Furthermore, the order in which the devices are considered also has a significant impact on efficiency, as it is shown in Ray2, Ray3 and Mandelbrot. When the slowest device processes the empty regions of these problems, its speedup is increased compared with other regions, unbalancing the execution.
The Dynamic algorithm has good results in most irregular applications when every device can provide enough computing capacity (Batel), achieving a geometric mean efficiency of 0.81, but suffers in benchmarks like NBody and Gaussian. They are sensitive to the number of chunks and their size, increasing the overhead of communication and usage of slow devices, respectively. Therefore, it is important to accurately determine the number of packages to get the best results in each benchmark. In Remo, the Dynamic algorithm suffers the penalization because of its weak CPU, which imbalances the co-execution when a wrong package size is giving to the slowest device.


Figure \ref{fig:worksizes} depicts the work size distribution between the devices for every scheduler configuration and benchmark. Each bar has three rectangles with the work size given to each device.
Every scheduling configuration distributes a similar workload for each device, except NBody and Mandelbrot, in Batel.
The CPU takes more workload as the number of packages increases in NBody, introducing smaller synchronization overheads when using fewer packages. Also, Mandelbrot shows how the Phi processed too much amount of work for the part of the image given in the Static, being more complex to calculate than the expected when computing the complete image. Also, Remo work distribution shows how the CPU penalized the whole execution in Dynamic due to large work sizes.

As it was introduced, the GPU in Binomial outperforms the CPU and Xeon Phi, as can be seen in the Static work size distributions. Therefore, a slightly variation in the completion time for any of these devices will imbalance the execution. Another important point is introduced to the analysis: the Xeon Phi's OpenCL driver needs and uses the CPU for its management. When using the CPU in co-execution, the Xeon Phi driver needs to share the CPU to build and manage the device with the CPU OpenCL driver, introducing time variations during the initialization and new overheads in the final completion times. This behavior is depicted in Figure \ref{fig:timingsBinomial}, on the left side, showing the average times from initialization for all the executions in Binomial, where the abscissa axis shows the base case (single device) and each scheduling configuration, with a bar showing the behavior for each device. The ordinate axis shows the time since EngineCL started. Using only the Phi needs around 1800 ms. to initialize and start computing, while it is up to 2700 ms. when using in Static. This variation combined with the small amount of work given to the CPU and Phi produces enough imbalance to not achieve the goal. On the other side, the Dynamic approach it is much worth for two reasons: it allows small periods of CPU time without computation (between chunks) to the Phi driver and thanks to its adaptability solves the initialization variations giving more chunks to the GPU, as it is shown. The drivers and its management are relevant to compute using OpenCL and produce an efficient co-execution, as can be seen on the right side of the figure, where Remo drivers and devices are completely stable compared with the Xeon Phi (Batel).

In summary, we can conclude that EngineCL can execute a single massive data-parallel kernel simultaneously on all devices in a heterogeneous system with a maximum overhead of 2.8\% and a tendency towards zero with bigger problem sizes. In addition, thanks to the load balancing algorithms, it yields excellent efficiencies, with different types of benchmarks. Therefore, this work has a multi-objective approach: offering high usability thanks to its API, while it has a flexible architecture that allows to extend its functionalities, offering low overheads in the worst case scenario, and is able to exploit all the devices of the heterogeneous system with high efficiency.

%% file: validation_usability_table_chart.tex

\begin{figure*}[!bp]
\captionof{table}{Comparison of Usability Metrics for a set of programs implemented in OpenCL and EngineCL (left). The figure shows the typical ratios found between both implementations (right). The programs have been selected based on the amount of properties, computing patterns and use cases to compare both implementations with enough diversity and using the worst-case scenario for EngineCL: using a single device.}
\vspace{-10mm}
\begin{minipage}{\textwidth}
\begin{minipage}[b]{0.69\textwidth}

  \label{tbl:usability}
\scriptsize
\begin{tabular}{p{14mm}l>{\columncolor[gray]{0.95}}rr>{\columncolor[gray]{0.95}}rr>{\columncolor[gray]{0.95}}rr>{\columncolor[gray]{0.95}}rr>{\columncolor[gray]{0.95}}rr>{\columncolor[gray]{0.95}}rr>{\columncolor[gray]{0.95}}rr>{\columncolor[gray]{0.95}}rr>{\columncolor[gray]{0.95}}rr}
\textbf{\textbf{Program}} & \textbf{\textbf{Runtime}} & \textbf{\textbf{CC}} & \textbf{\textbf{TOK}} & \textbf{\textbf{OAC}} & \textbf{\textbf{IS}} & \textbf{\textbf{LOC}} & \textbf{\textbf{INST}} & \textbf{\textbf{MET}} & \textbf{\textbf{ERRC}}\\
\hline

Gaussian   & OpenCL               & 4   & 585  & 312 & 433 & 87 & 17 & 28 & 22 \\
           & EngineCL             & 1   & 60   & 33  & 53  & 15 & 3 & 13 & 1 \\
           & \ \ \ \ \ \ \ ratio  & 4 & 9.8  & 9.5 & 8.2 & 5.8 & 5.7 & 2.2 & 22.0 \\ \hline
Ray        & OpenCL               & 4   & 618  & 307 & 424 & 89 & 17 & 27 & 21 \\
           & EngineCL             & 1   & 191  & 40  & 65  & 24  & 3 & 17 & 1 \\
           & \ \ \ \ \ \ \ ratio  & 4 & 3.2  & 7.7 & 6.5 & 3.7 & 5.7 & 1.6 & 21.0 \\ \hline
Binomial   & OpenCL               & 4   & 522  & 255 & 355 & 77 & 16 & 24 & 18 \\
           & EngineCL             & 1   & 81   & 28  & 48  & 18 & 3 & 11 & 1 \\
           & \ \ \ \ \ \ \ ratio  & 4 & 6.4  & 9.1 & 7.4 & 4.3 & 5.3 & 2.2 & 18.0 \\ \hline
Mandelbrot & OpenCL               & 4   & 473  & 222 & 313 & 74 & 15 & 24 & 18 \\
           & EngineCL             & 1   & 65   & 35  & 55  & 15 & 3 & 13 & 1 \\
           & \ \ \ \ \ \ \ ratio  & 4 & 7.3  & 6.3 & 5.7 & 4.9 & 5 & 1.8 & 18.0 \\ \hline
NBody      & OpenCL               & 4   & 658  & 373 & 517 & 96 & 18 & 32 & 26 \\
           & EngineCL             & 1   & 66   & 38  & 60  & 16 & 3 & 15 & 1 \\
           & \ \ \ \ \ \ \ ratio  & 4 & 10.0 & 9.8 & 8.6 & 6.0 & 6.0 & 2.1 & 26.0 \\ \hline
           & \ \ \ \ \ \rule{0pt}{3ex}\boldmath$\overline{ratio}$  & \textbf{4:1} & \textbf{7.3} & \textbf{8.5} & \textbf{7.3} & \textbf{4.9} & \textbf{5.5} & \textbf{2.0} & \textbf{21.0} \\

\end{tabular}

\end{minipage}
\hspace{-7mm}
\begin{minipage}[b]{0.31\textwidth}
  \vspace{1cm}
   \centering
\raisebox{-0.5\height}{
   \includegraphics[width=1.09\textwidth]{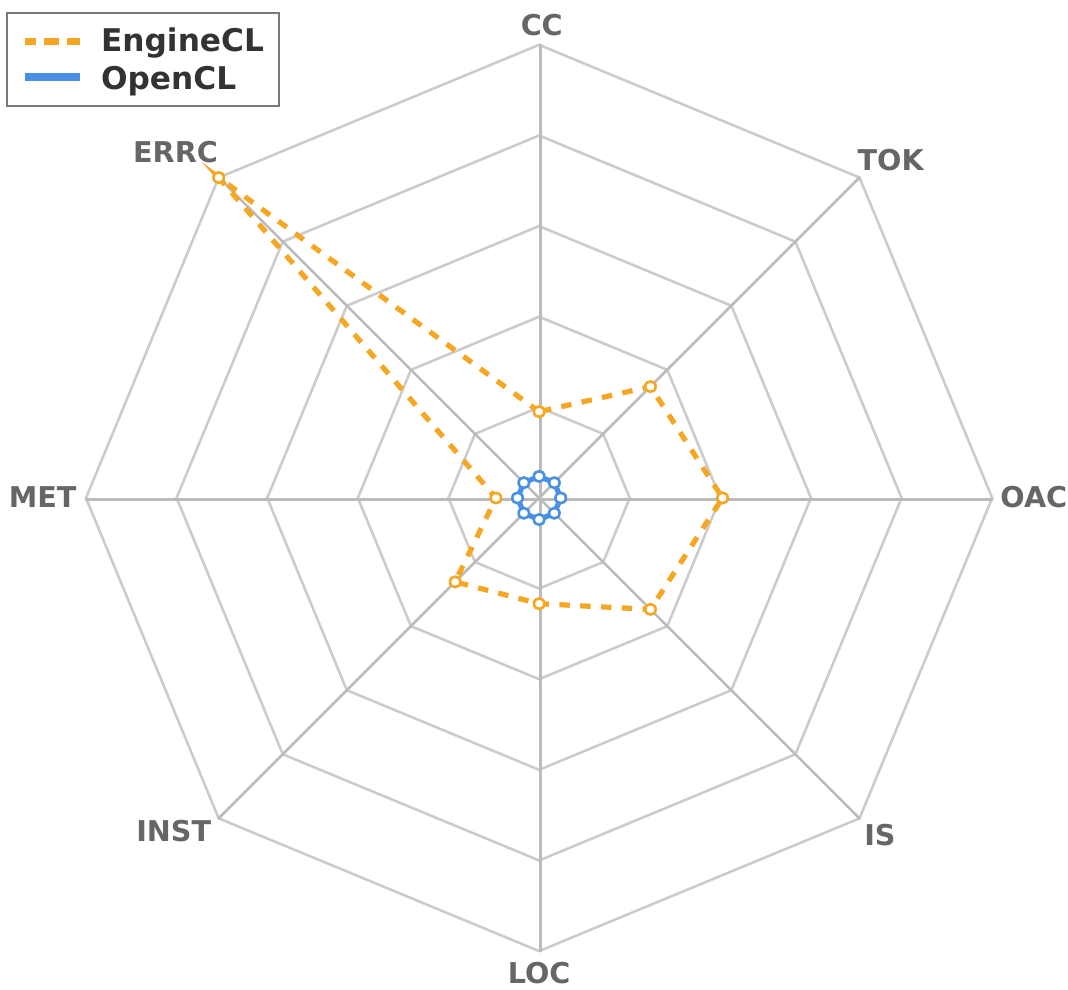}
}

\end{minipage}
\end{minipage}
\end{figure*}

%% file: related.tex
\section{Related Work}
\label{sec:Related}

There are projects aiming at high-level parallel programming in \Cpp{}, but most of them provide an API similar to the Standard Template Library (STL) to ease the parallel programming, like Boost.Compute \cite{Boost.Compute:2016}, HPX \cite{HPX:2017}, Thrust \cite{Thrust:2009}, SYCL \cite{SYCL:2017} and the \Cpp{} Extensions for Parallelism TS \cite{ParallelSTL:2015}.
Thrust, tied to CUDA devices, HPX, which extends the C++ Parallel and Concurrency TS with asynchronous algorithm and additional future types for distributed computing, or \Cpp{} TS are not OpenCL-centered.
Projects like HPX.Compute \cite{HPX.Compute:2017} and SYCLParallelSTL \cite{SyclParallelSTL:2015} provide backends for OpenCL via SYCL. 
SYCLParallelSTL exposes ParallelSTL on CPU and GPU.
Proposals like SkelCL \cite{SkelCL:2011} and SkePU \cite{SkePU:2010}
provide data management and composable primitives and skeletons to build parallel applications,
but the programmer is responsible of using their own data containers.
EngineCL offers a high-level layered API with better usability than the previous \Cpp{} proposals, generally provide constructs based on STL.
Also, there are C-programmed libraries with similar objectives, but they provide low-level APIs where the programmer needs to specify many parameters and the density of the code is considerable. While Maat \cite{Maat:2009} uses OpenCL to achieve the code portability, Multi-Controllers \cite{MultiControllers:2017} is CUDA and OpenMP-centered, but allows kernel specialization. On the other side, EngineCL targets a 
flexible
API with an application domain as execution unit, increasing significantly the productivity. It provides different API layers, allows kernel specialization, direct usage of \Cpp{} containers, manages the data and work distribution transparently between devices and has smaller overheads compared with the previous projects.

On the other hand, there is a considerable amount of bibliography relating to task parallelism. In this case the workload is independent tasks belonging to the same application, which are distributed to different devices. For instance, \cite{dongarra}, proposes a lightweight runtime based on QUARK, that uses a greedy heuristic. The authors of \cite{Zhang} apply fuzzy neural networks to the task distribution problem. MultiCL \cite{pena} is an OpenCL runtime based on storing execution information for each kernel-device pair for future kernel launches. VirtCL is a framework based on OpenCL \cite{You} which constructs regression models to predict task turnaround times. SPARTA \cite{Donya:2016} analyzes tasks at runtime and uses the obtained information to schedule the next tasks maximizing energy-efficiency. Finally, Unicorn \cite{unicorn} and Xkaapi\cite{xkaapi} are parallel programming models based on a work-stealing task scheduler.

Regarding load balancing algorithms some approaches use a static algorithm, like the implemented in EngineCL. As instance, Kim \emph{et al.} \cite{achieving} implements an OpenCL framework that provides the programmer with a view of a single compute, but only consider systems with several identical GPUs and ignore the CPUs, so their proposal is not suitable for truly heterogeneous systems. Lee \emph{et al.}\cite{lee} propose the automatic modification of OpenCL code that executes on a single device, so the load is balanced among several ones. De la Lama \emph{et al.} \cite{lbcl} propose a library that encapsulates standard OpenCL calls. The works presented in \cite{Kofler} and \cite{Zhong} use machine learning and performance models respectively, to come up with an offline model that predicts an ideal static load partitioning, but it does not consider irregularity. \cite{Lee2} modifies the code to get a static distribution of a single kernel to the available devices. Other authors have proposed training-based methods to the load balancing problem, like Qilin \cite{qilin} and Maestro \cite{maestro}.

Finally, there are several works that uses dynamic load balancing algorithms. FluidicCL \cite{fluidic} implements an adaptive dynamic scheduler, but only focuses on systems with one CPU and one GPU. SnuCL \cite{snucl} is an OpenCL framework for heterogeneous CPU/GPU clusters. However, it does not support the cooperative execution of a kernel using all the available devices. Kaleem \emph{et al.} in \cite{adapt} and Boyer \emph{et al.} in \cite{Skadron} propose adaptive methods that use the execution time of the first packages to distribute the remaining load. However, they focus on a CPU/GPU scenario and do not scale well to configurations with more devices. Similarly, HDSS \cite{hdss} dynamically learns the computational speed of each device during an adaptive phase and then schedules the remainder of the workload using a weighted self-scheduling scheme during the completion phase. However, this algorithm  assumes lineal speed with the package size, which might not be true for irregular kernels. Similarly, a dynamic and adaptive algorithm for TBB is proposed in \cite{Asenjo, asenjo_logfit}. This is also based on using small initial packages to identify a package size that obtains near optimal performance. Finally, Finepar \cite{finepar} builds a performance model to come up with the ideal work partition for an irregular application. For this approach, the performance model calculation needs to be performed every time the input data changes, which is costly.


%% file: conclusions.tex
\section{Conclusions and Future Work}
\label{sec:Conclusions}

Over the last decade heterogeneous systems have become ubiquitous in a wide family of computing devices, from high performance computing nodes, to desktop computers and smartphones, thanks to their excellent performance, power consumption and energy efficiency. But this heterogeneity and diversity of devices pose major challenges to the community. This paper addresses some of these challenges, such as the complexity of programming and performance portability, and it proposes effortless  co-execution as one of the key concepts to overcome them.

For this purpose, EngineCL is presented, a powerful OpenCL-based tool that greatly simplifies the programming of applications for heterogeneous systems. This runtime frees the programmer from tasks that require a specific knowledge of the underlying architecture, and that are very error prone, with a great impact on their productivity. Moreover, the runtime is designed and profiled to provide internal flexibility to support new features, high performance to avoid any overheads compared with OpenCL and a pluggable scheduling system to efficiently use all the available resources with custom load balancers. The API provided to the programmer is very simple, thus improving the usability of heterogeneous systems. Besides, it also ensures performance portability thanks to the integration of schedulers that successfully distribute the workload among the devices, adapting both to the heterogeneity of the system and to the behavior of the applications.

These statements are corroborated by the exhaustive validation that is presented, both in usability and performance. Regarding usability, a large variety of well-known and widely used Software Engineering metrics has been analyzed, achieving excellent results in all of them. The performance has been validated in two different nodes, one HPC and one commodity system, with six different architectures to show the compatibility and efficiency of EngineCL. Three important conclusions can be drawn. First, the careful design and implementation of EngineCL allows
small overheads with respect to the native OpenCL version, always below 2.8\% in all the cases studied and with an average overhead of 1.3\% considering the worst-case scenario. Furthermore, EngineCL scales very well with the size of the problem, so overheads vanish for large problem sizes. Second, it is critical to select the right scheduler, especially for irregular applications, where it needs to be dynamic and adaptive. Among the schedulers implemented and integrated in EngineCL, HGuided provides the best results, being able to balance both regular and irregular applications, with an average efficiency of 0.89 and 0.82 for the HPC and desktop system, respectively. Finally, thanks to all the above, EngineCL is able to provide the programmer with effortless co-execution, thus ensuring performance portability between very different heterogeneous systems.


In the future, it is intended to extend the API to support iterative and multi-kernel executions. Also, new load balancing algorithms will be provided and studied as part of the scheduling system, focusing on performance and energy efficiency.

%% file: biblio.tex
\bibliographystyle{splncs03}

%% file: bio.tex
\pagebreak
\begin{description}
\hyphenation{Can-ta-bria}
\InsertBoxL{0}{%
  \includegraphics[width=25mm,height=25mm,keepaspectratio,clip]{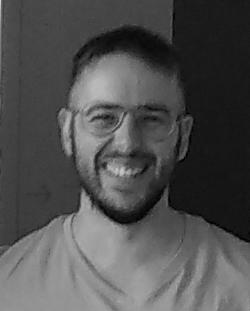}%
  \vspace{5mm}%
}[0]
{\small \textbf{Ra\'ul Nozal}
is a Computer Engineer and Topography Engineer, expert in Software Architecture (Graduated in UPV/EHU, UNICAN and UDIMA Universities). He has professional background and mentored some business projects and R\&D software. Currently pursuing a PhD degree on Technology and Science in the Universidad de Cantabria. He is a Pre-PhD researcher in the Dept. of Computer Engineering and Electronics. His research interests include programming languages, heterogeneous systems, HPC, web technologies and software architecture.}

\vspace{3mm}
\InsertBoxL{0}{\includegraphics[width=25mm,height=25mm,keepaspectratio]{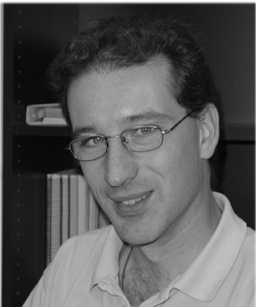}}[-1]
{\small \textbf{Jose Luis Bosque} graduated in Computer Science from Universidad Polit\'ecnica de Madrid in 1994. He received the PhD degree in Computer Science and Engineering in 2003 and the Extraordinary Ph.D Award from the same University. He joined the Universidad de Cantabria in 2006, where he is currently Associate Professor in the Department of Computer and Electronics. His research interests include high performance computing, heterogeneous systems and interconnection networks.}

\vspace{3mm}
\hyphenation{Ca-ta-lun-ya}
\InsertBoxL{0}{\includegraphics[width=25mm,height=25mm,keepaspectratio]{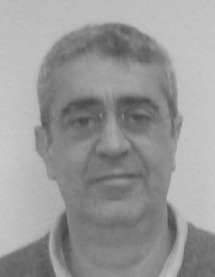}}[-1]
{\small \textbf{Ram\'on Beivide} received the B.Sc. and M.Sc. degrees in Computer Science from the Universidad Aut\'onoma de Barcelona (UAB) in 1981 and 1982. The Ph.D. degree, also in Computer Science, from the Universidad Polit\'ecnica de Catalunya (UPC) in 1985.  He joined the Universidad de Cantabria in 1991, where he is currently a Professor in the Dept. of Computer Engineering and Electronics. His research interests include computer architecture, interconnection networks, coding theory and graph theory.
}
\item
\end{description}